\documentclass{cernrep}
\usepackage{makecell}
\usepackage{fontsize}
\usepackage{lineno}
\usepackage{hyphenat}
\newlength\cmsTabSkip

\begin{document}

\title{Experimental road of the $J/\psi \phi$ mass spectrum, current status, and implications}
\author{Hongjian Zhou$^{1}$, Xining Wang$^{2}$\footnote{Both are the first author. Contact emails: Kai.yi@njnu.edu.cn}, Feng Zhu$^{1}$, Liming Zhang$^{3}$, Gerry Bauer$^{1}$, and Kai Yi$^{1}$}

\institute{Ministry of Education Key Laboratory of NSLSCS, Institute of Physics Frontiers and Interdisciplinary Sciences \\ $^1\text{School of Physics and Technology, Nanjing Normal University, Nanjing 210023, China}$ \\
$^2\text{Department of Physics, Tsinghua University, Beijing 100084, China}$ \\
$^3\text{Department of Engineering Physics, Tsinghua University, Beijing 100084, China}$}

\begin{abstract}

Inspired by the $X(4140)$ structure reported in the $J/\psi \phi$ system by the CDF experiment in 2009, a series of searches have been carried out in the $J/\psi \phi$ and $J/\psi K$ channels, leading to the claim of ten structures in the $B \rightarrow J/\psi \phi K$ system. This article provides a comprehensive review of experimental developments, from the initial evidence of $X(4140)$ at CDF to the amplitude analyses and diffractive process investigations by the LHCb experiment, as well as theoretical interpretations of these states. 
A triplet of $J^{PC} = 1^{++}$ states with relatively large mass splittings [about 200~MeV (natural units are adopted)] has been identified in the $J/\psi \phi$ system by LHCb. Their mass-squared values align approximately linearly with a possible radial quantum number, suggesting that the triplet may represent a radially excited family. 
For $X(4140)$, the first state in the triplet, its width reported by LHCb is inconsistent with that measured by other experiments, and possible reasons for this discrepancy are discussed.
A potential connection between an excess at 4.35~GeV in the $J/\psi \phi$ mass spectrum reported by the Belle experiment through a two-photon process and a potential spin-2 excess in the LHCb data is also investigated.
In addition, potential parallels between the $J/\psi \phi$ and $J/\psi J/\psi$ systems, both composed of two vector mesons, are discussed. The continued interest in, and complexity of, these systems underscore the necessity of further experimental exploration with increased statistical precision across a variety of experiments, particularly those with relatively flat efficiency across the Dalitz plot. The $J/\psi \omega$, $\phi \phi$, $\rho \omega$, and $\rho \phi$ systems are mentioned, and the prospects for the $J/\psi \Upsilon$ and $\Upsilon \Upsilon$ systems, are also highlighted.

\end{abstract}


\maketitle

%%%%%%%%%%%%%%%%%%%%%%%%%%%%%%%%%%%%%%%%%%%%%%%%%%%%%%%%%%%%%%%%%%%%%%%%%%%%
\section{Introduction}
\indent

Heavy-flavor quarks have historically played a vital role in establishing and validating the quark model, most notably exemplified by the discovery and study of the $J/\psi$ meson and its excitations. Similarly, they have been central to the exploration and identification of exotic hadronic states, as extensively reviewed in~\cite{Zhu:2024swp}. Among the quarks, charm ($c$), bottom ($b$), and top ($t$) are classified as heavy flavors; up ($u$) and down ($d$) as light flavors; while strange ($s$) occupies an intermediate position. While systems composed solely of heavy quarks, such as $X \rightarrow J/\psi J/\psi$~\cite{Zhu:2024swp}, provide a unique perspective on exotic hadrons, the present review concentrates on systems without light quarks, with emphasis on the $J/\psi \phi$ system.

Experimental reviews of structures in the $J/\psi \phi$ mass spectrum have been published in 2013~\cite{YI:2013iok} and 2018~\cite{Yi_2018}. Since then, several new structures have been discovered~\cite{LHCb:2021uow,LHCb_2024_diffractive,LHCb_2023_B0_JpsiKs0}. With more than a decade having passed since the first review, substantial progress has been made in studies of the $J/\psi \phi$ system, as well as in the development of the $J/\psi K^{+}$ system. The exploration of the $J/\psi \phi$ system has proven complex and challenging, and its history continues to unfold. Recent results from LHCb, using an amplitude analysis~\cite{LHCb:2021uow}, have revealed numerous new resonances, two of which have been independently confirmed. Yet the intricacies of this system remain unresolved, and much of its nature is still unknown. At present, the leading experiments investigating the $J/\psi \phi$ system are the CMS and LHCb experiments, both of which benefit from the significantly higher integrated luminosities collected in recent years. In light of these advances, a new comprehensive review of the subject is timely.

Progress has been made in the similar system \(J/\psi J/\psi\)~\cite{CMS:2023owd,ATLAS:2023bft,LHCb:2020bwg}, with a recent and significant advance achieved by CMS using data collected from 2016 to 2018 and from 2022 to 2024. Three structures,  named $X(6600)$, $X(6900)$, and $X(7100)$, are identified, and interference effects between them are determined. All these structures, together with their interferences, are observed with statistical significances well above 5$\sigma$, implying common $J^{PC}$ quantum numbers~\cite{JJRun3PAS}. $X(6900)$ has been confirmed, and evidence for $X(7100)$ has been reported in an independent decay mode, $J/\psi\psi(2S) \rightarrow \mu^{+}\mu^{-}\mu^{+}\mu^{-}$~\cite{JPRun3PAS}. The squared masses of the structures fall on a line consistent with radial excitations~\cite{Zhu:2024swp}, with widths that systematically decrease as the mass increases. CMS also measured their $J^{PC}$ using the 2016--2018 dataset, showing that the data favor the $2^{++}$ assignment, with other possibilities excluded at least at the 95\% confidence level~\cite{CMS:2025fpt}. These features disfavor molecular or threshold interpretations and instead support a $J^{PC} = 2^{++}$ tetraquark picture composed of aligned spin-1 diquarks without orbital excitation~\cite{JJRun3PAS,CMS:2025fpt}.

Spin-2 states are rare among both conventional and exotic hadrons. Since 2003, when the first heavy-exotic was discovered~\cite{CDF:2003cab}, around 60 exotic hadrons have been discovered, including 23 at the Large Hadron Collider (LHC), yet none have been firmly established with \(J=2\). The \(J/\psi \phi\) and \(J/\psi J/\psi\) systems provide favorable environments to search for spin-2 particles, as both the $J/\psi$ and the $\phi$ mesons carry spin-1. However, the production of spin-2 particles can be suppressed if they originate from a spin-0 system, such as the $B$ meson, as will be discussed in detail later. A potential structure with $J^{P}=2^{-}$ has been reported in the \(J/\psi \phi\) system~\cite{LHCb:2021uow}, but it does not reach the 5$\sigma$ significance threshold and has not been confirmed by other experiments. Another possible spin-2 state, around the 4.35~GeV region in the $J/\psi \phi$ mass spectrum, is noted in this article. Among the many states claimed in the \(J/\psi \phi\) system, three share identical quantum numbers, $J^{PC} = 1^{++}$, yet display large mass splittings of about 200~MeV. These states may potentially form a family of radial excitations.

This article begins with a brief recap of studies in the $J/\psi \phi$ system prior to 2013, and then primarily outlines the experimental landscape of the $J/\psi \phi$ structures from 2013 to the most recent developments. The main focus is on the $J/\psi \phi$ system, while the $J/\psi K$ system is also briefly discussed, since these two can correspond to different two-body subsystems within the same three-body decay process $B^+ \to J/\psi \phi K^+$, and are dynamically coupled.
These experimental results are combined with theoretical studies to shed further light on the nature of exotic hadrons. In addition, the connection between the $J/\psi J/\psi$ and $J/\psi \phi$ systems is discussed, and other related vector-vector and hidden-flavor systems similar to the $J/\psi \phi$ system are also briefly mentioned for comparison.

\section{The rich structures in the $J/\psi \phi K$ system}
\subsection{Experimental facilities}
\indent

The $J/\psi \phi$ system has been studied across a wide range of facilities, including hadron collider experiments (CDF, CMS, $\mathrm{D\textsl{\O}}$, and LHCb), and $e^+e^-$ collider experiments (BaBar, Belle, and BESIII). The $e^+e^-$ machines provide a clean experimental environment with relatively low background levels, while the hadron collider experiments benefit from significantly higher production cross sections, allowing them to access rare decay modes and higher-mass regions. Each experiment thus offers complementary strengths in the search for exotic structures. Basic information about these detectors and their relevant features is summarized in Tab.~\ref{tab:detector}.

\begin{table}[htbp!]
    \centering
    \newcolumntype{Y}{>{\raggedleft\arraybackslash}X}
    \caption{Summary of the experimental detectors that are relevant to the study of structures in the $J/\psi \phi$ system. In the column $\sqrt{s}$, values in parentheses (such as $8.0/3.5$) indicate the individual beam energies, where the first number corresponds to the electron energy in~GeV and the second to the positron energy in~GeV. 
    The colliders include the Tevatron at Fermilab, 
    the KEK B-Factory (KEKB) at the High Energy Accelerator Research Organization (KEK),
    the Electron-Positron Project II (PEP-II) at the Stanford Linear Accelerator Center (SLAC), 
    the Beijing Electron–Positron Collider II (BEPCII) at the Institute of High Energy Physics (IHEP), 
    and the LHC at the European Organization for Nuclear Research (CERN).
    }
    \begin{tabular}{@{}cccccc@{}} 
    \hline
    Accelerator&Detectors&Type&$\sqrt{s}$ [GeV] & Running years & Laboratory\\
    \hline
    Tevatron & CDF~\cite{CDF_detector_1988}, $\mathrm{D\textsl{\O}}$~\cite{D0:1993bqa} & $p\bar{p}$&\makecell[c]{$1800$ (Run 1)\\$1960$ (Run 2)}&\makecell[c]{$1992-1996$  \\ $2001-2011$ } & Fermilab\\
    KEKB & Belle~\cite{belle_detector} & $e^+e^-$ & $10.58$ ($3.5/8.0$) & $1998-2010$ &KEK\\
    Super-KEKB & Belle II~\cite{Belle-II:2010dht} & $e^+e^-$ & $10.58$ ($4.0/7.0$) & $2018-\cdots$ & KEK\\
    PEP-II & BaBar~\cite{BaBar_detector_2001} & $e^+e^-$ & $10.58$ ($3.1/9.0$) & $1999-2008$ & SLAC\\
    BEPCII & BESIII~\cite{BESIII_detector}&$e^{+}e^{-}$ & $2.0-4.95$ & $2008-\cdots$ & IHEP\\
    LHC&\makecell[c]{CMS~\cite{CMS_detector}, LHCb~\cite{LHCb_detector}} & $pp$ & \makecell[c]{ $>7000$ (Run 1)\\ $13000$ (Run 2)\\ $13600$ (Run 3)} & \makecell[c]{$2009-2013$ \\ $2015-2018$ \\ $2022-\cdots$} & CERN\\
    \hline
    \end{tabular}
    \label{tab:detector}   
\end{table}

\subsection{Structures in $J/\psi \phi$ system before LHCb's amplitude analyses}
\indent

\begin{figure}[!htbp]
    \centering
    \includegraphics[width=0.29\linewidth]{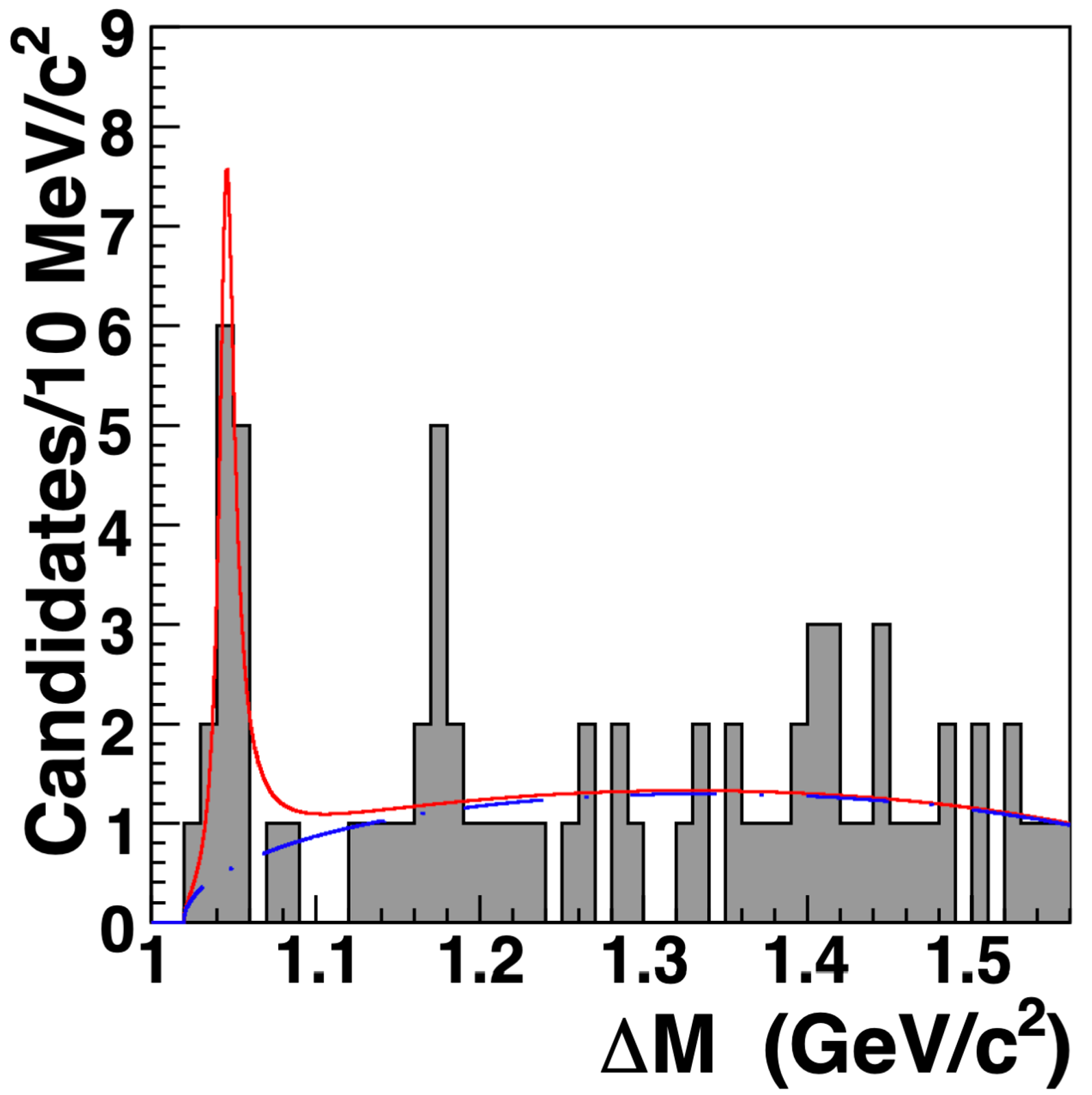}
    \includegraphics[width=0.3\linewidth]{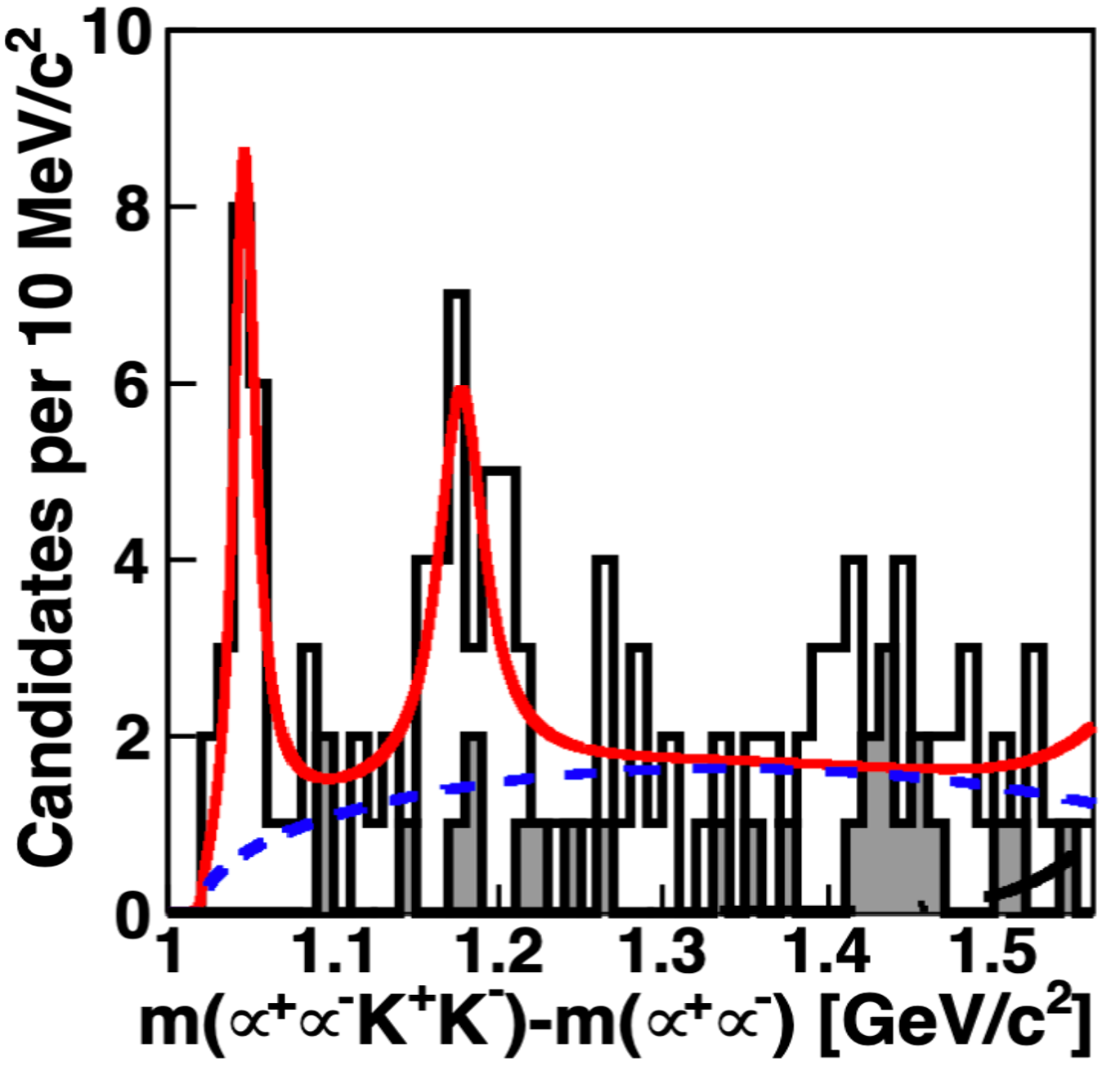}
    \includegraphics[width=0.38\linewidth]{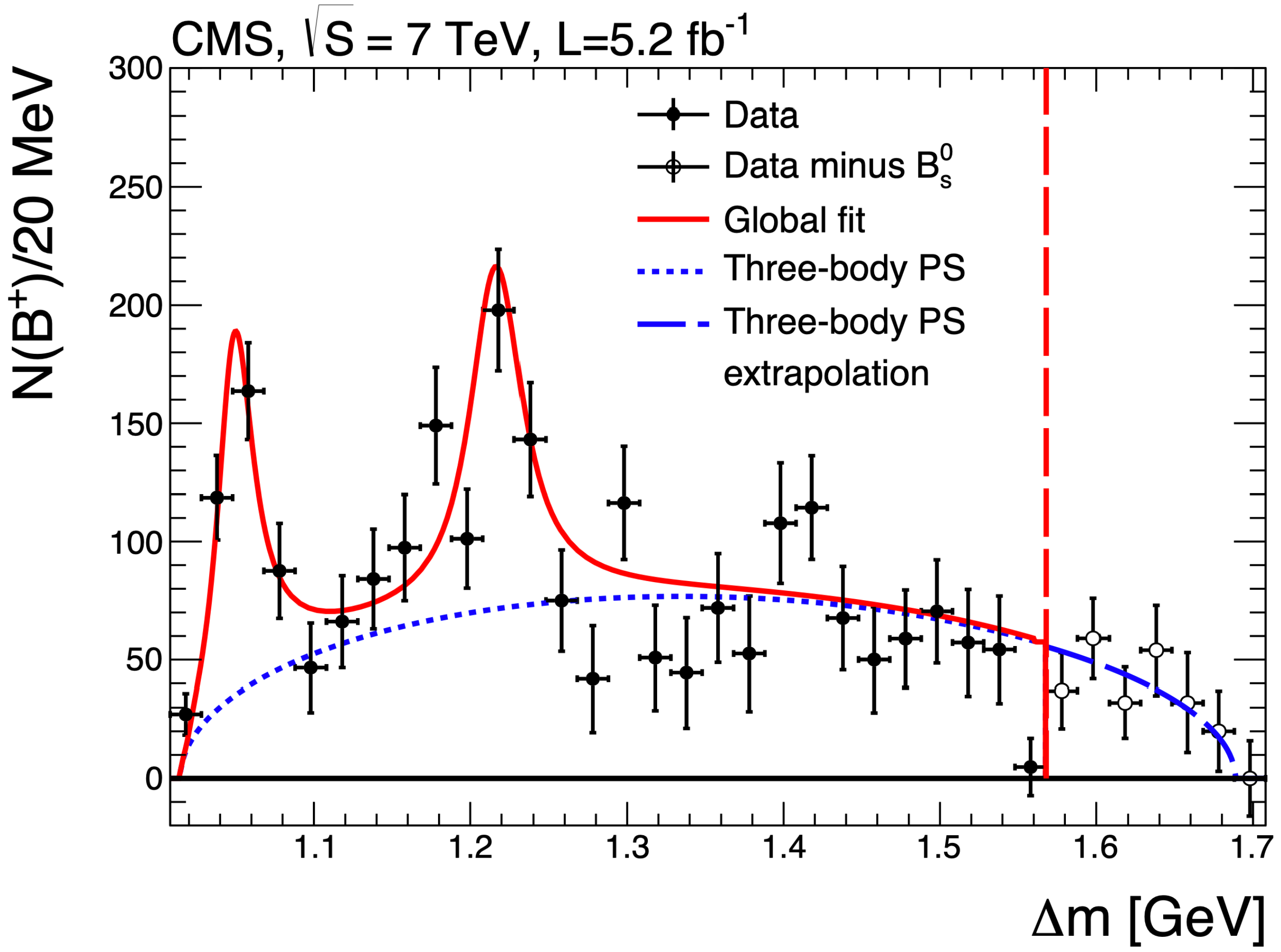}
    \caption{The number of $B^{+} \rightarrow J/\psi \phi K^{+}$ candidates as a function of the mass difference between $\mu^{+}\mu^{-}K^{+}K^{-}$ and $\mu^{+}\mu^{-}$ from CDF in 2009~\cite{CDF:2009jgo} (left),  2011~\cite{CDF:2011pep} (middle), and the results from CMS in 2013~\cite{CMS:2013jru} (right).}
    \label{fig:CDF_Y4140}
\end{figure}

There has been considerable interest in the $J/\psi\phi$ mass spectrum in $B^{+}\rightarrow J/\psi \phi K^{+}$ decays since the report of $X(4140)$ by CDF in 2009 (Fig.~\ref{fig:CDF_Y4140}, left)~\cite{CDF:2009jgo}. This state, originally dubbed $Y(4140)$ and now cataloged as $\chi_{c1}(4140)$~\cite{ParticleDataGroup:2024cfk} (here we adopt the notation $X$ for many particles in this paper), was reported with a statistical significance of $3.8\sigma$. Its mass was $4143.0 \pm 2.9  \pm 1.2 $~MeV and its width was $11.7^{+8.3}_{-5.0}\pm3.7$~MeV (the first uncertainty is statistical and the second is systematic throughout the paper). Unlike $X(3872)$, the first heavy exotic candidate which was discovered in the $J/\psi \pi^{+} \pi^{-}$ spectrum~\cite{Belle:2003nnu}, $X(4140)$ features a distinct quark composition, notably incorporating the heavier strange quark. 

\begin{figure}[!htbp]
    \centering
    \includegraphics[width=0.48\linewidth]{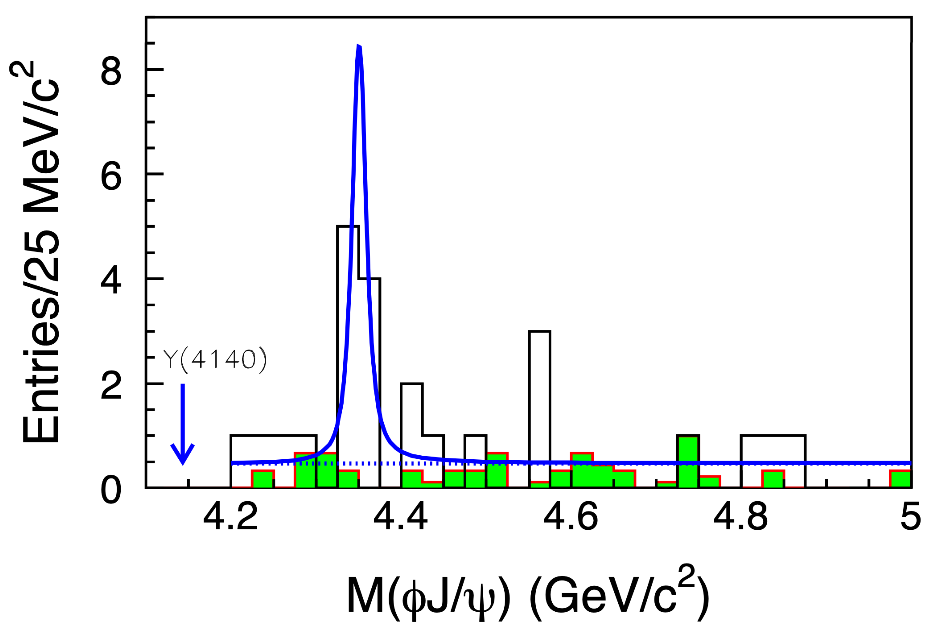}
    \includegraphics[width=0.48\linewidth]{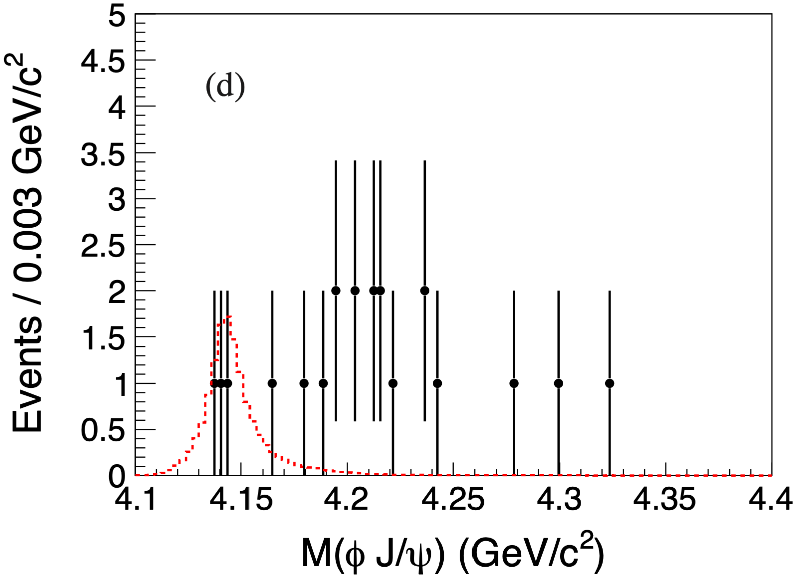}
    \caption{Left: The \(J/\psi \phi\) invariant-mass distribution from Belle, with the data represented by the open histogram. The solid line corresponds to the fit model, while the dashed line describes the background. The expected position of $X(4140)$ is indicated by the blue arrow~\cite{Belle:2009rkh}. Right: The mass distribution of $J/\psi \phi$ from BESIII, with a red dashed line representing signal Monte Carlo (MC) events of $e^{+}e^{-}\rightarrow \gamma X(4140)$, $X(4140) \rightarrow J/\psi \phi$~\cite{BESIII:2014fob}.}
    \label{fig:Belle_BESIII_noY4140}
\end{figure}

In 2010, Belle investigated the two-photon process $\gamma \gamma \rightarrow J/\psi \phi$ and found no signal corresponding to $X(4140)$. This result posed a challenge to the interpretation of $X(4140)$ as a $D_s^{*+} D_s^{*-}$ molecular state with quantum numbers \(J^{PC} = 0^{++}\) or \(2^{++}\)~\cite{Belle:2009rkh}. Instead, the experiment reported evidence for another narrow structure, $X(4350)$, with a significance of 3.2$\sigma$, as illustrated in the left panel of Fig.~\ref{fig:Belle_BESIII_noY4140}. 
In 2011, with a larger dataset, CDF confirmed the existence of $X(4140)$ with a significance exceeding $5\sigma$ (mass was $4143.4^{+2.9}_{-3.0}\pm0.6$~MeV and width was $15.3^{+10.4}_{-6.1}\pm2.5$~MeV). Furthermore, a second structure with a mass of $4274.4 ^{+8.4}_{-6.7} \pm 1.9$~MeV and a width of $32.3^{+21.9}_{-15.3}\pm7.6$~MeV was identified with a significance of $3.1\sigma$ (Fig.~\ref{fig:CDF_Y4140} middle)~\cite{CDF:2011pep}. 
In 2012, LHCb did not confirm $X(4140)$, while not excluding its existence~\cite{LHCb:2012wyi}. This discrepancy delayed the publication of the 2011 CDF result, which was eventually published in 2017~\cite{CDF:2011pep} after the confirmation by CMS~\cite{CMS:2013jru}. 

\begin{figure}[!htbp]
    \centering
    \includegraphics[width=0.43\linewidth]{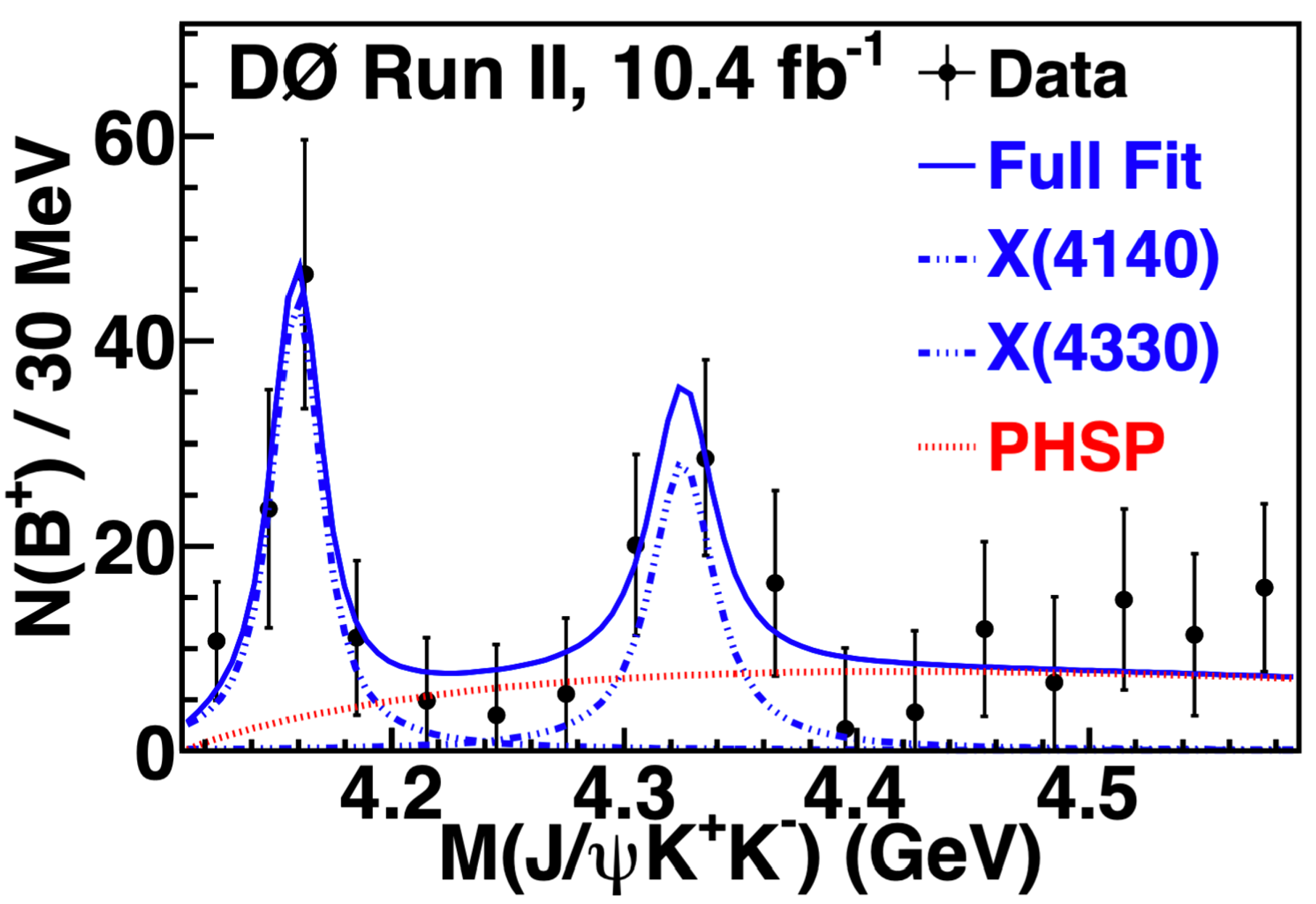}
    \includegraphics[width=0.39\linewidth]{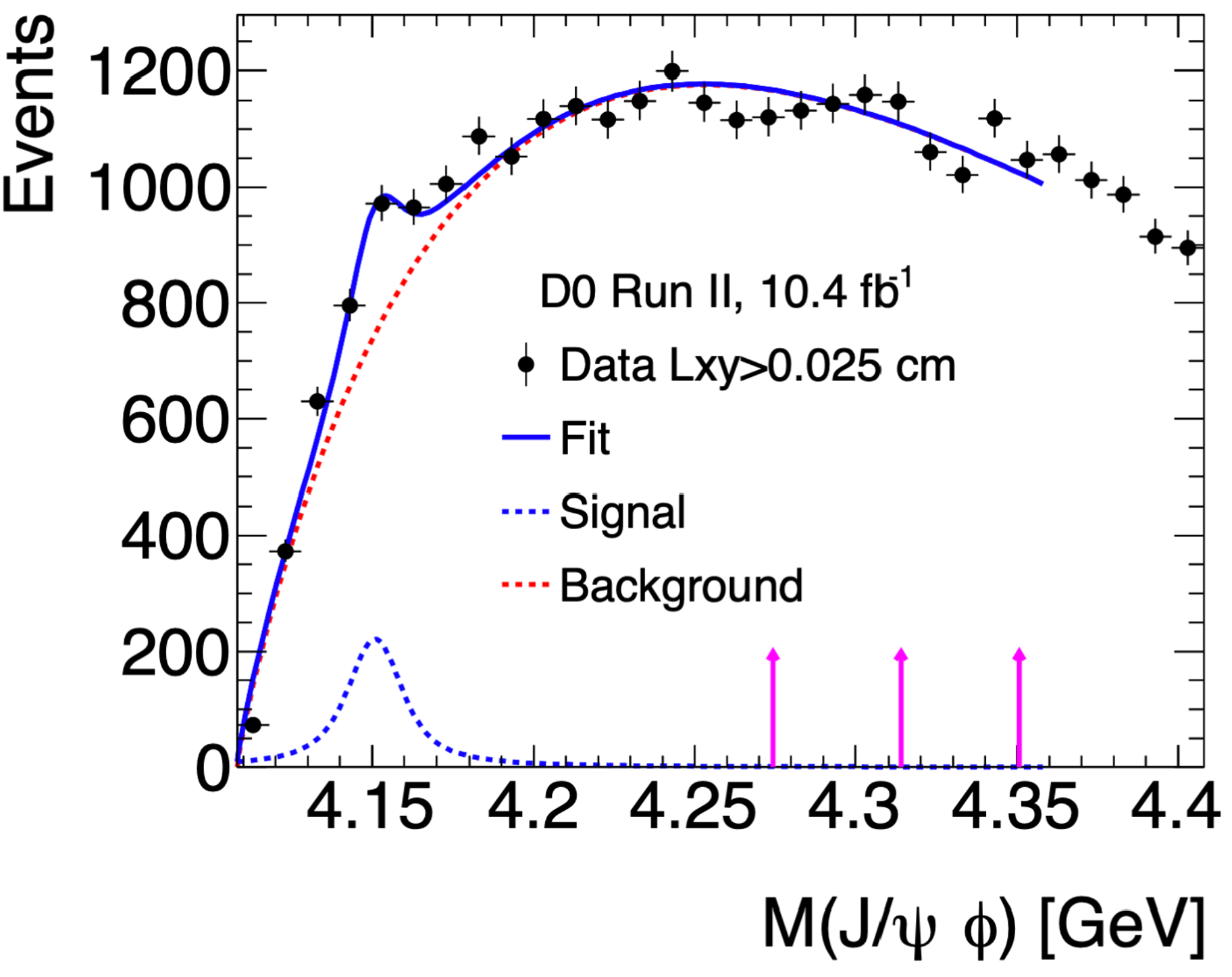}
    \caption{Invariant-mass distribution of $J/\psi \phi$ candidates through the $B^{+} \rightarrow J/\psi \phi K^{+}$ decay channel~\cite{D0:2013jvp} (left) and through inclusive production~\cite{D0:2015nxw} (right) from $\mathrm{D\textsl{\O}}$. In the left panel, the width of the second Breit-Wigner resonance is fixed to 30~MeV (consistent with CDF data~\cite{CDF:2009jgo}).}
    \label{fig:D0_Y4140}
\end{figure}

\begin{figure}[!htbp]
    \centering
    \includegraphics[width=0.48\linewidth]{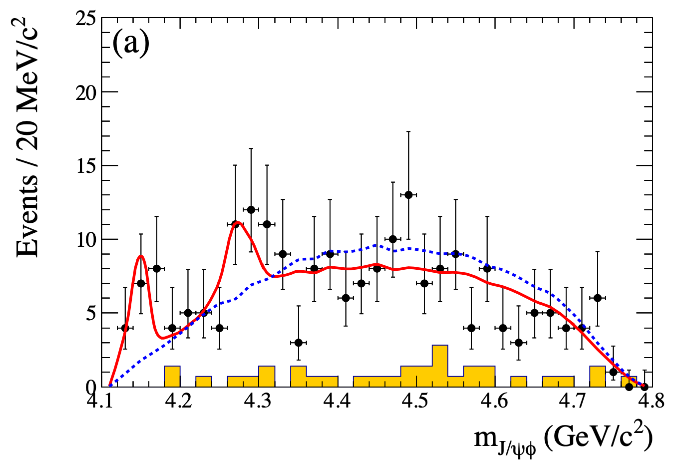}
    \includegraphics[width=0.48\linewidth]{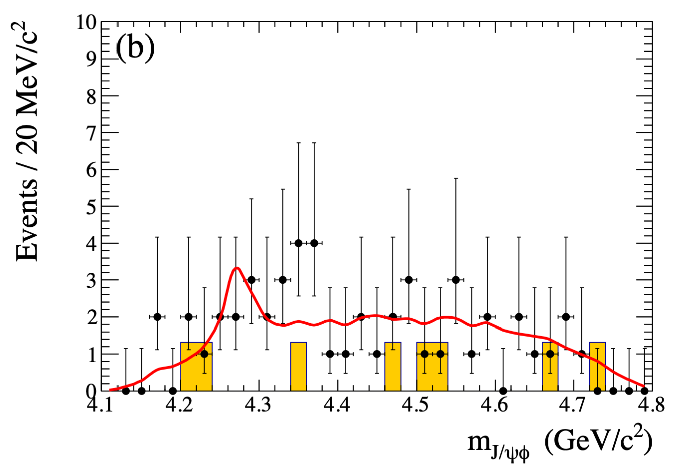}
    \caption{The $J/\psi \phi$ invariant-mass distributions for $B^{+}\rightarrow J/\psi \phi K^{+}$ (left) and $B^{0} \rightarrow J/\psi \phi K_{S}^{0}$ (right) from BABAR. The projections from the Dalitz plot fit, including $X(4140)$ and $X(4272)$ (solid red lines) are shown ($X(4140)$ is not clearly visible for $B^{0}$ data sample). A fit without resonances (dashed blue line) is also shown in the left panel. The shaded yellow histograms represent the estimated background, determined from the $\Delta E$ sideband regions~\cite{BaBar:2014wwp}.}
    \label{fig:BABAR_Y4140}
\end{figure}

In 2003, CMS confirmed the existence of $X(4140)$~\cite{CMS:2013jru}, 
% and provided evidence for an additional structure at $4313.8 \pm 5.3 \pm 7.3$ MeV, with a width of $38^{+30}_{-15} \pm 7.3 $ MeV in 2013~\cite{CMS:2013jru}, 
as depicted in Fig.~\ref{fig:CDF_Y4140} (right). Shortly thereafter, $\mathrm{D\textsl{\O}}$ reported a structure in $B^{+} \rightarrow J/\psi \phi K^{+}$ decays, identifying a mass of $4159.0\pm 4.3\pm 6.6$~MeV and a width of $19.9 \pm 12.6^{+1.0}_{-8.0} $~MeV, with a statistical significance of $3.1\sigma$, which was consistent with $X(4140)$ reported by CDF. Additionally, the data showed a hint of a possible second state at $4328.5 \pm 12.0\,\mathrm{MeV}$~\cite{D0:2013jvp}, as illustrated in Fig.~\ref{fig:D0_Y4140} (left). 
In 2014, BESIII examined the decay process $e^{+}e^{-} \rightarrow \gamma \phi J/\psi$ at $\sqrt{s}=4.23$, $4.26$, and $4.36$~GeV, but no significant $J/\psi \phi$ structure was observed~\cite{BESIII:2014fob}, as indicated in the right graph of Fig.~\ref{fig:Belle_BESIII_noY4140}. During the same year, BABAR explored the decay $B^{\pm,0} \rightarrow J/\psi \phi K^{\pm,0}$ and found no evidence for resonance in the $J/\psi \phi$ mass spectrum~\cite{BaBar:2014wwp}, as presented in Fig.~\ref{fig:BABAR_Y4140}.
In 2015, $\mathrm{D\textsl{\O}}$ further examined the inclusive $J/\psi \phi$ samples via the $p\bar{p} \rightarrow J/\psi \phi + {anything}$ process and reported a structure at $4152.5 \pm 1.7 ^{+6.2}_{-5.4}\,\mathrm{MeV}$, with a width of $16.3 \pm 5.6 \pm 11.4$~MeV and a significance of $4.7\sigma$~\cite{D0:2015nxw}, as shown in Fig.~\ref{fig:D0_Y4140} (right).
The observation of $X(4140)$ was reported by some experiments but not confirmed by others, which may be attributed to the limited statistical precision of the datasets and differences in detector performance. Nevertheless, the mass and width measurements of the $X(4140)$ reported by these experiments were generally consistent within their uncertainties.

\subsection{Latest results with LHCb amplitude analyses}
\indent

\renewcommand{\arraystretch}{1.3}
\begin{table}[h]
\scriptsize
    \centering
    \caption{Structures reported in the $J/\psi\phi$ mass spectra from different experiments. The average width of $X(4140)$ is calculated using 2011 CDF, 2013 CMS, 2013 $\mathrm{D\textsl{\O}}$, and 2015 $\mathrm{D\textsl{\O}}$ results. The significances, evaluated with statistical uncertainties only, are marked with $'*'$.}
    \begin{tabular}{ccccccccc} 
    \hline
    \hline
    Year & Experiment & Luminosity & Process/Yield & Structure & Mass  &  Width & $J^{P}$ &Significance  \\
    & & [$\mathrm{fb}^{-1}$] & $B\rightarrow J/\psi \phi K$ & & [$\mathrm{MeV}$] & [$\mathrm{MeV}$] & & [$\sigma$] \\
    \hline
    2009 & CDF~\cite{CDF:2009jgo} & 2.7  & $75\pm10$ & $X(4140)$ & $4143.0\pm2.9\pm1.2$ & $11.7^{+8.3}_{-5.0}\pm3.7$ &  & $3.8^{*}$  \\
    2010 & Belle~\cite{Belle:2009rkh} & 825  & $\gamma \gamma \rightarrow J/\psi \phi$ & $X(4350)$ & $4350.6^{+4.6}_{-5.1}\pm0.7$ & $13^{+18}_{-9}\pm4$ &  & $3.2$  \\
    2011 & CDF~\cite{CDF:2011pep} &$6.0$ &  $115\pm12$ & $X(4140)$ & $4143.4^{+2.9}_{-3.0}\pm0.6$ & $15.3^{+10.4}_{-6.1}\pm2.5$ & & $>5.0^{*}$ \\ 
    & & &  & $X(4274)$ & $4274.4^{+8.4}_{-6.7} \pm 1.9$  & $32.3^{+21.9}_{-15.3}\pm7.6 $ & & $3.1^{*}$ \\
    2012 & LHCb~\cite{LHCb:2012wyi}& 0.37  & $346\pm20$ & $X(4140)$ & 4143.0 (fixed) & 15.3 (fixed) \\
    2013 & CMS~\cite{CMS:2013jru}& $5.2$  & $2480\pm160$ & $X(4140)$ & $4148.0\pm2.4\pm6.3$ & $28^{+15}_{-11}\pm19$ &  & $>5.0^{*}$  \\
    2013 & $\mathrm{D\textsl{\O}}$~\cite{D0:2013jvp}& 10.4  & $215\pm37$ & $X(4140)$ & $4159.0\pm4.3\pm6.6$ & $19.9\pm12.6^{+1.0}_{-8.0}$ &  & $3.1^{*}$\\
    2014 & BaBar~\cite{BaBar:2014wwp}& $422.5$  & $189\pm14$ & $X(4140)$ & 4143.4 (fixed) & 15.3 (fixed) &  & $1.6$  \\
    2014 & BESIII~\cite{BESIII:2014fob}&  &$e^{+}e^{-} \rightarrow \gamma J/\psi \phi$ & -- & -- &-- & &-- \\
    2015 & $\mathrm{D\textsl{\O}}$~\cite{D0:2015nxw}& $10.4$ &$p\bar{p} \to J/\psi \phi + \text{anything}$ & $X(4140)$ & $4152.5\pm1.7^{+6.2}_{-5.4}$ & $16.3\pm 5.6\pm 11.4$ &  & $4.7$  \\
    \hline
    Average&&&& $X(4140)$ &$4146.8\pm 2.4$&$17.3^{+6.6}_{-5.3}$ &&  \\
    \hline
    \hline
    2016 & LHCb~\cite{LHCb:2016axx} & 3 & $4289\pm 151$& $X(4140)$ & $4146.5\pm4.5^{+4.6}_{-2.8}$ & $83\pm21^{+21}_{-14}$ & $1^{+}$ & $8.4$\\
    &&&&$X(4274)$ &$4273.3\pm8.3^{+17.2}_{-3.6}$ & $56\pm11^{+8}_{-11}$ & $1^{+}$ &  $6.0^{*}$ \\
    &&&&$X(4500)$ &$4506\pm11 ^{+12}_{-15}$ &  $92\pm21^{+21}_{-20}$ & $0^{+}$ & $6.1^{*}$ \\
    &&&&$X(4700)$ &$4704\pm10^{+14}_{-24}$ & $120\pm31^{+42}_{-33}$ & $0^{+}$ &  $5.6^{*}$ \\
    \hline
    2021 & LHCb~\cite{LHCb:2021uow}& 9& $24220\pm 170$ & $X(4140)$ & $4118 \pm 11^{+19}_{-36}$ & $162 \pm 21^{+24}_{-49}$ & $1^{+}$ & 13 \\
    &&&& $X(4150)$ &$4146 \pm 18 \pm 33$ & $135 \pm 28 ^{+59}_{-30}$& $2^{-}$ & 4.8\\
    &&&& $X(4274)$ & $4294 \pm 4 ^{+3}_{-6}$ & $53 \pm 5 \pm5$& $1^{+}$ & 18\\
    &&&& $X(4500)$ & $4474 \pm 3 \pm 3 $ &$77 \pm 6^{+10}_{-8}$ & $0^{+}$ &20\\
    &&&& $X(4630)$ & $4626 \pm 16 ^{+18}_{-110}$ & $174 \pm 27 ^{+134}_{-73}$& $1^{-}$ & 5.5\\
    &&&& $X(4685)$ & $4684 \pm 7^{+13}_{-16}$ &$126 \pm 15^{+37}_{-41}$& $1^{+}$ & 15 \\
    &&&& $X(4700)$ & $4694 \pm 4^{+16}_{-3}$ &$87 \pm 8^{+16}_{-6}$& $0^{+}$ & 17 \\
    \hline
    2024 & LHCb~\cite{LHCb_2024_diffractive}& 5& $pp\rightarrow J/\psi \phi + \text{anything} $ & $X(4274)$ & $4298 \pm 6 \pm 9 $ & $92^{+22}_{-18}\pm57$ & & 4.1 \\
    &&&& $X(4500)$ & $4512.5^{+6.0}_{-6.2}\pm 3.0$ & $65 ^{+20}_{-16}\pm 32$& &6.1\\
    \hline
    \hline
    \end{tabular}
    \label{tab:jpsiphi_mass_spectrum}
\end{table}

The latest results on the $J/\psi \phi$ system were reported by LHCb using the amplitude fits. 
Unlike earlier studies based solely on invariant-mass fits, which could not disentangle overlapping enhancements or determine quantum numbers, the amplitude analysis exploited the multidimensional kinematic information of the decay. 
Simultaneously fitting the invariant masses and decay angles of the studied systems accounted for quantum interferences among different intermediate states. 
Through the multidimensional likelihood fit, each resonance is characterized by its mass, width, and quantum numbers, with the spin and parity determined by testing alternative amplitude hypotheses and comparing their likelihood values.
This multidimensional approach enabled the extraction of the spin–parity quantum numbers, dealt with reflections that could mimic signal structures, and resolved multiple overlapping resonances that were indistinguishable in one-dimensional fits.

In 2016, LHCb conducted the first amplitude analysis of $B^{+} \rightarrow J/\psi \phi K^{+}$ decays, reestablishing the existence of $X(4140)$ and $X(4274)$~\cite{LHCb:2016axx}.
The width of $X(4140)$ was found to be significantly larger than previously determined, as detailed in Tab.~\ref{tab:jpsiphi_mass_spectrum}. Both structures were identified as $1^{+}$ resonances. Additionally, the high $J/\psi\phi$ mass region was explored for the first time with good sensitivity, uncovering two significant structures interpreted as $0^{++}$ resonances: $X(4500)$ and $X(4700)$.

\begin{figure}
    \centering
    \includegraphics[width=1\linewidth]{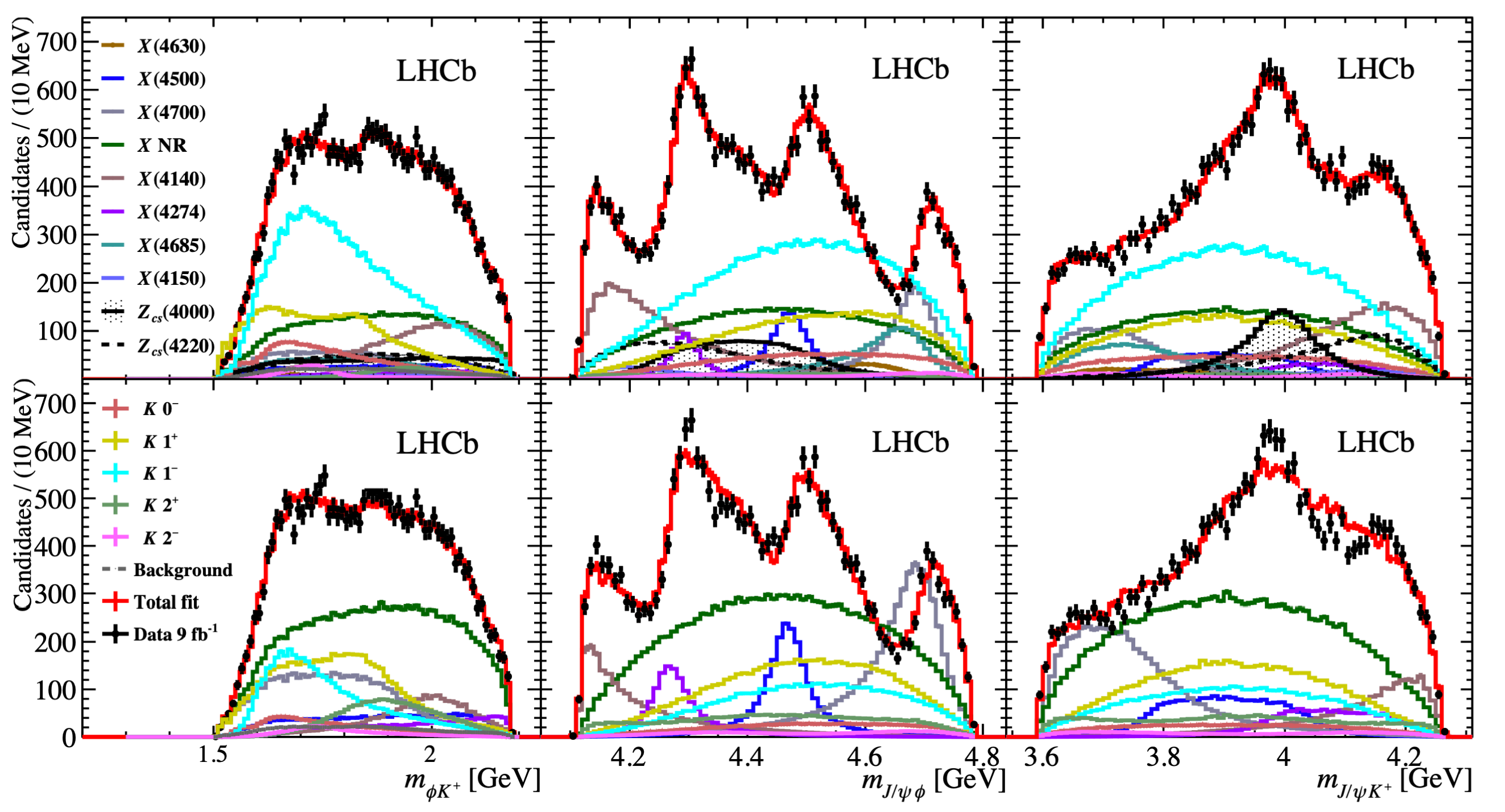}
    \caption{Invariant-mass distributions of $\phi K^{+}$ (left), $J/\psi \phi$ (middle), and $J/\psi K^{+}$ (right) for the $B^{+} \rightarrow J/\psi \phi K^{+}$ candidates in LHCb's Run 1 and Run 2 data, shown together with the LHCb 2021 model (top) and the LHCb 2016 model (bottom)~\cite{LHCb:2021uow}. Compared to the LHCb 2016 model, the main change in the LHCb 2021 model is to introduce additional states, such that no further state with statistical significance above 5$\sigma$ needs to be added to improve the fit to the combined Run 1 and Run 2 data. For the LHCb 2021 model, nine $\phi K^{+}$ (states with same spin parity are merged for display and labeled as $K$ in the legend), seven $J/\psi \phi$ (labeled as $X$), two $J/\psi K^{+}$ (labeled as $Z_{cs}$) and one $J/\psi \phi$ nonresonant (labeled as $X$ NR) components are considered.}
    \label{fig:LHCb2021_Jpsiphi_massSpetrum}
\end{figure}

In 2021, LHCb made substantial improvements in the full amplitude analysis, achieving a signal yield ($24220\pm170$) for the combined Run 1 and Run 2 data approximately six times higher than in previous studies in 2016.
The LHCb 2016 model was tested but failed to provide satisfactory description of data (Fig.~\ref{fig:LHCb2021_Jpsiphi_massSpetrum}, bottom). The increased statistics required an improved fit model. A new model (LHCb 2021 model), was introduced by adding more components until no additional contribution with a statistical significance greater than $5\sigma$ was found. Specifically, in the LHCb 2021 model, all components from the LHCb 2016 model were retained, while the main change was the inclusion of additional $J/\psi \phi$ and $J/\psi K$ states [$X(4150)$, $X(4685)$, $X(4630)$, $Z_{cs}(4000)^{+}$, and $Z_{cs}(4220)^{+}$, to be introduced later]. The fit result is presented in Fig.~\ref{fig:LHCb2021_Jpsiphi_massSpetrum} (top).
The study reaffirmed the existence of previously observed four $J/\psi \phi$ states with greater significance and confirmed their quantum number assignments. Additionally, a new $J^{P} = 1^{+}$ state, designated as $X(4685)$, was identified with a 
high significance ($15\sigma$). A state, $X(4630)$, was revealed with a significance of 5.5$\sigma$. This state favors a spin-parity assignment of $1^{-}$ over $2^{-}$ at the 3$\sigma$ level and excludes other $J^{P}$ hypotheses at the 5$\sigma$ level. Furthermore, a $J^{P}=2^{-}$ particle $X(4150)$ was reported with a significance of $4.8\sigma$ (statistical significance is $8.7\sigma$). In total, seven structures were reported in the $J/\psi \phi$ mass spectrum in this analysis.

In addition to the $J/\psi \phi$ structures, LHCb unveiled a relatively narrow state, $Z_{cs}(4000)^+$ (also known as $T_{c\bar{c} \bar{s} 1}(4000)^{+}$~\cite{ParticleDataGroup:2024cfk}, while we adopt the notation $Z_{cs}$ to refer to $J/\psi K$ structure), decaying to $J/\psi K^+$ with a mass of $4003 \pm 6 ^{+4}_{-14}$~MeV and a width of $131 \pm 15\pm26$~MeV~\cite{LHCb:2021uow}. The spin parity of this state was determined to be $1^+$, and the phase motion observed in the amplitude confirmed its resonance nature. Furthermore, the analysis revealed the necessity of a broader $1^+$ or $1^-$ $Z_{cs}(4220)^{+}$ at a significance level of 5.9$\sigma$. This analysis marks the first observation of states with hidden charm and naked strangeness decaying to the $J/\psi K^+$ final state.

\begin{figure}[!htbp]
    \centering
    \includegraphics[width=0.98\linewidth]{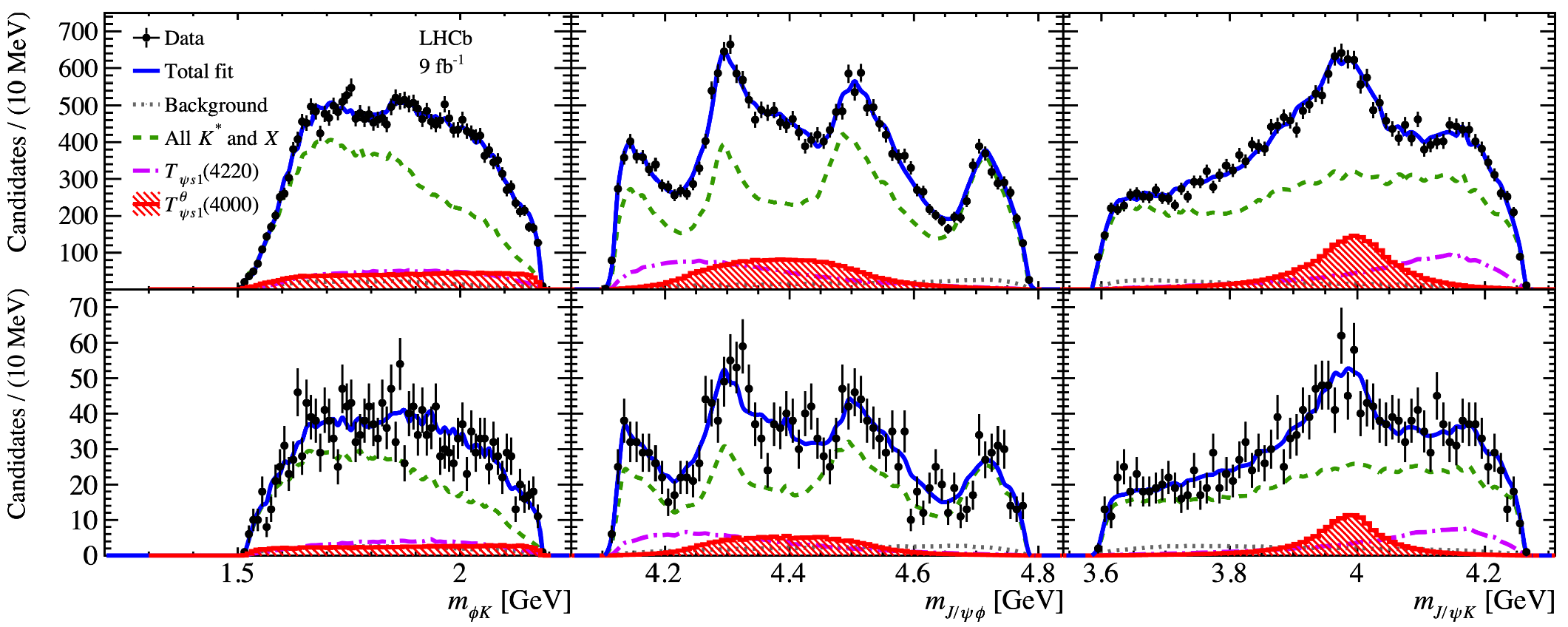}
    \caption{Invariant-mass distribution from LHCb of $\phi K$ (left), $J/\psi \phi$ (middle) and $J/\psi K$ (right) with the projections of the LHCb 2021 model for $B^{+} \rightarrow J/\psi \phi K^{+}$ (top) and $B^{0} \rightarrow J/\psi \phi K^{0}_{S}$ (bottom)~\cite{LHCb_2023_B0_JpsiKs0}. The data and model in the upper plots are the same ones as those in the upper panels of Fig.~\ref{fig:LHCb2021_Jpsiphi_massSpetrum}, but with all $\phi K$ and $J/\psi\phi$ structures merged together to highlight the $J/\psi K$ structures (green dashed curve), $Z_{cs}(4000)$ and $Z_{cs}(4220)$ (labeled as $T_{\psi s1}$ in the legend). For the fit to the $B^{0}$ sample, the parameters of all components except for $Z_{cs}(4000)^{0}$ are taken to be the same as those obtained from the $B^{+}$ sample, under the assumption of isospin symmetry. The parameters of $Z_{cs}(4000)^{0}$ in $B^{0}$ decay are treated independently from those of $Z_{cs}(4000)^{+}$ in $B^{+}$ decay. The parameters of $Z_{cs}(4220)^0$ state are fixed to the values from $Z_{cs}(4220)^{+}$ owing to the limited statistics of the $B^{0}$ sample.}
    \label{fig:2023_LHCb_B0_JpsiKs0HCb}
\end{figure}

In 2023, LHCb investigated the decay \(B^{0} \to J/\psi \phi K^{0}_{S}\) to analyze the \(J/\psi K^{0}_{S}\) mass spectrum~\cite{LHCb_2023_B0_JpsiKs0}. Evidence was found for the $Z_{cs}(4000)^{0}$ state at a significance of \(4\sigma\), suggesting it may be the isospin partner of $Z_{cs}(4000)^{+}$. Under the assumption of isospin symmetry, the significance of $Z_{cs}(4000)^{0}$ increases to \(5.4\sigma\). These $J/\psi K$ structures are illustrated in Fig.~\ref{fig:2023_LHCb_B0_JpsiKs0HCb} and Tab.~\ref{tab:jpsiK_mass_spectrum}.

\renewcommand{\arraystretch}{1.3}
\begin{table}[!htbp]
\scriptsize
    \centering
    \caption{Structures reported in the $J/\psi K$ mass spectrum by LHCb.}
    \begin{tabular}{cccccccc} 
    \hline
    \hline
    Year & Experiment & Luminosity & Process & Structure & Mass  &  Width & Significance  \\
    & & [$\mathrm{fb}^{-1}$] & $B\rightarrow J/\psi \phi K$ & & [$\mathrm{MeV}$] & [$\mathrm{MeV}$] & [$\sigma$] \\
    \hline
    2021 & LHCb~\cite{LHCb:2021uow} & 9 & $B^{+} \rightarrow J/\psi \phi K^{+}$ & $Z_{cs}(4000)^{+}$ & $4003\pm 6 ^{+4}_{-14}$ & $131\pm15\pm26$  & $15.0$  \\
    &&&& $Z_{cs}(4220)^{+}$ & $4216 \pm 24 ^{+43}_{-30}$ & $233\pm 52^{+97}_{-73}$ &  $5.9$ \\
    2023 & LHCb~\cite{LHCb_2023_B0_JpsiKs0} & 9 & $B^{0} \rightarrow J/\psi \phi K^{0}_{S}$ &$Z_{cs}(4000)^{0}$ & $3991^{+12+9}_{-10-17}$ & $105^{+29+17}_{-25-23}$  & $4.0$  \\
    \hline
    \hline
    \end{tabular}
    \label{tab:jpsiK_mass_spectrum}
\end{table}

\begin{figure}[!htbp]
    \centering
    \includegraphics[width=0.5\linewidth]{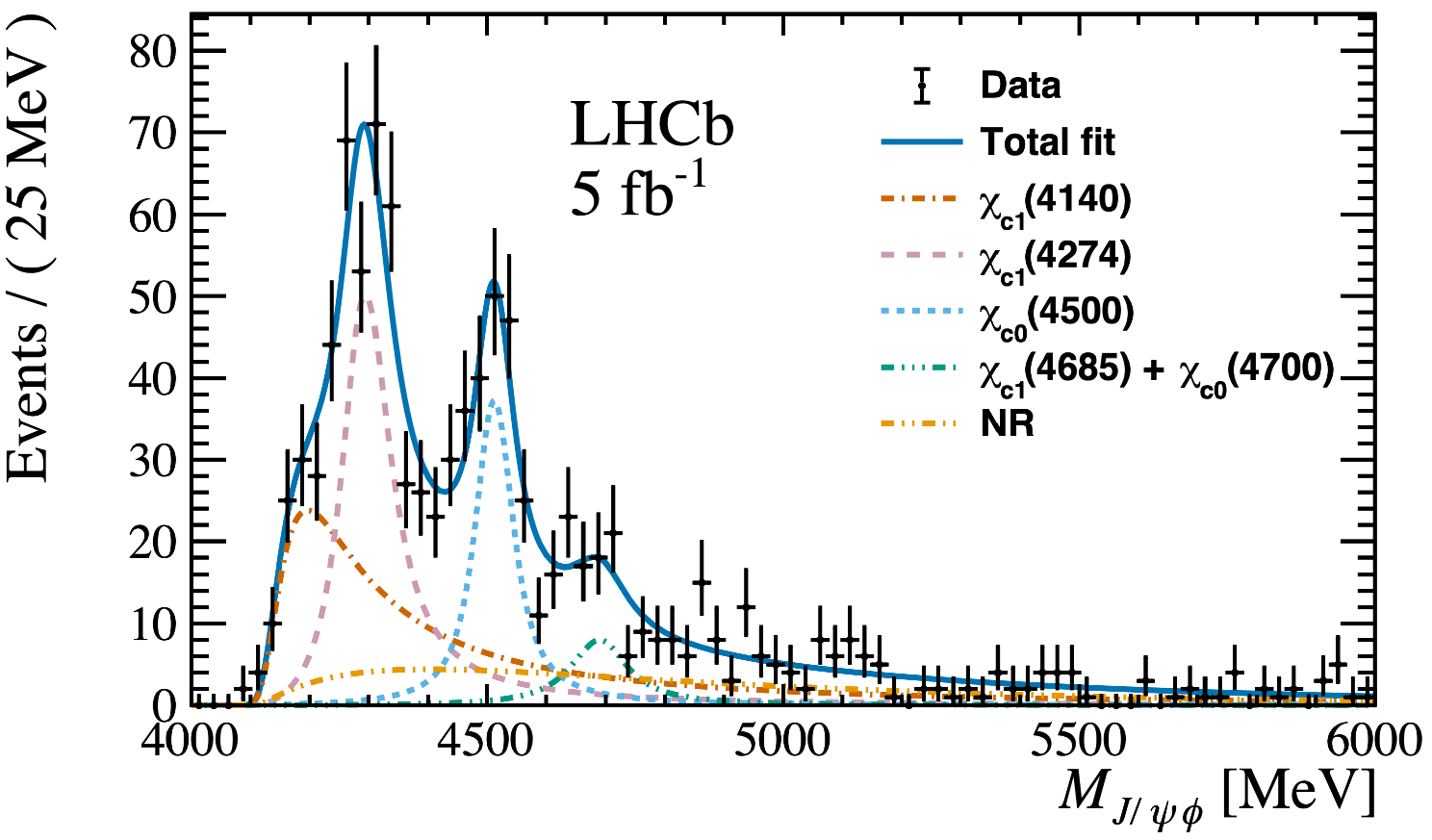}
    \caption{The mass spectrum of $J/\psi \phi$ from diffractive processes in proton-proton collisions from LHCb~\cite{LHCb_2024_diffractive}. In the fit model, there are five structures, $X(4140)$, $X(4274)$, $X(4500)$, $X(4685)$, and $X(4700)$ (the label $\chi_{c1}$, rather than $X$, is used in the legend). The mass and width of weaker structures [$X(4140)$, $X(4685)$, and $X(4700)$] are fixed to the LHCb 2021 results~\cite{LHCb:2021uow} due to the limited statistics. 
    % The significances of $X(4274)$ and $X(4500)$ are evaluated.
    }
    \label{fig:2024_LHCb_diffractvie}
\end{figure}

In 2024, the first study of $J/\psi \phi$ production in diffractive processes in proton-proton collisions was conducted by LHCb~\cite{LHCb_2024_diffractive}. In such processes, the protons typically remain or dissociate only slightly, and the events are characterized by low multiplicity and large rapidity gaps devoid of particles, where the suppressed hadronic activity may enhance the sensitivity to rare produced states. The results were consistent with the previously claimed structures~\cite{LHCb:2021uow}: $X(4500)$ was observed with a significance exceeding 5\(\sigma\), and the $X(4274)$ was confirmed with a significance above 4\(\sigma\) (Fig.~\ref{fig:2024_LHCb_diffractvie}).

There is a rich spectrum of exotic candidates involving both $c$ and $s$ quarks. In total, ten new $J/\psi \phi$ and $J/\psi K$ structures have been claimed in $B \rightarrow J/\psi \phi K$ decays by LHCb. However, apart from the $X(4140)$ and $X(4274)$, these states have not yet been reported by other experiments. Additionally, the width of the $X(4140)$ reported by LHCb is inconsistent with that measured by other experiments, warranting further investigation.

\section{Theoretical interpretations}
\indent

Various theoretical models have been proposed to explain the observed structures in the $J/\psi \phi$ and $J/\psi K$ invariant-mass spectra. However, no model has been definitively established. These models can be broadly classified into conventional charmonium states, compact tetraquarks, molecular states, hybrid configurations, and nonresonant effects.

%%%%%%%%%%%%%%%%%%%%%%%%%%%%%%%%%%%%%%%%%%%%%%%%%%%%%%%%%%%%%%%%

\begin{itemize}
    \item {\bfseries Conventional charmonium interpretations:}
Some of the observed structures have been interpreted as conventional excited charmonium states within the framework of potential models. For example, $X(4140)$ has been proposed as a candidate for the $\chi_{c1}(3P)$ $c\bar{c}$ state~\cite{Hao:2019fjg}, while $X(4274)$ and $X(4500)$ have been associated with other P-wave charmonium states~\cite{Ortega:2016hde,Bokade:2024tge}.
    \item {\bfseries Compact tetraquark interpretations:}
Compact tetraquark models, which are typically composed of diquark-anti\-diquark pairs~\cite{Chen:2022asf}, provide an alternative explanation. A diquark is a correlated pair of two quarks weakly bound together inside a hadron~\cite{Barabanov:2020jvn,GellMann}. For instance, $X(4140)$ and $X(4274)$ both have been interpreted as D-wave $cs\bar{c}\bar{s}$ tetraquark states with $J^{P}=1^{+}$~\cite{Chen:2016oma}. Numerous other studies have assigned various $J/\psi \phi$ structures to tetraquark configurations~\cite{Lu:2016cwr,Wang:2016tzr,Wu:2016gas,Wang:2016gxp,Maiani:2016wlq,Deng:2017xlb,Yang:2019dxd}.
    \item {\bfseries Molecular interpretations:}
Hadronic molecules, i.e., meson-meson bound states, represent another plausible explanation. 
For example, $Z_{cs}(4000)$ has been proposed as a pure molecular state, expressed as $\frac{1}{\sqrt{2}} \left( \left| \bar{D}_s^* D \right\rangle \pm \left| \bar{D}_s D^* \right\rangle \right)$~\cite{MENG20212065}. Many other heavy-quark molecular states have also been predicted~\cite{Huang:2024asn}.
    \item {\bfseries Hybrid interpretations:}
Hybrid hadrons, in which quarks are bound together with a valence gluon~\cite{Wan:2020fsk}, i.e., $q\bar{q}g$, have also been explored. For example, significant meson-hybrid admixtures have been claimed for $X(4140)$ and $X(4274)$ using QCD Laplace sum-rules~\cite{Palameta:2018yce}.
    \item {\bfseries Nonresonant interpretations:}
An alternative perspective considers the observed enhancements as arising from dynamic effects near thresholds, rather than from genuine resonances. Threshold cusps~\cite{Guo:2019twa}, resulting from the opening of new decay channels, can produce sharp features in invariant-mass distributions which can mimic peaks. Additionally, triangle singularities~\cite{Guo:2019twa}, occurring when intermediate particles within a loop diagram simultaneously go on shell, can generate localized enhancements that mimic resonant behavior. For instance, $Z_{cs}(4000)$, $Z_{cs}(4220)$, and $X(4700)$ have been interpreted as the $J/\psi K^{*+}$, $\psi(2S)K^{+}$, and $\psi(2S) \phi$ threshold cusps, respectively, with these cusps amplified by nearby triangle singularities~\cite{Ge:2021sdq}. Several theoretical studies have also employed threshold cusp mechanisms to account for various structures observed in the $J/\psi \phi$ and $J/\psi K$ systems~\cite{Liu:2016onn, Ikeno:2021mcb, Luo:2022xjx, Nakamura:2023swt}. 
These nonresonant mechanisms are claimed to reproduce key features of the data without introducing new particle states.
\end{itemize}

The theoretical landscape for interpreting the $J/\psi \phi$ and $J/\psi K$ structures is rich and multifaceted. While conventional charmonium assignments offer one avenue, the peculiar properties of the observed states have necessitated the exploration of exotic configurations and nonresonant effects. Distinguishing among these possibilities requires detailed amplitude analyses, high-statistics data, and a comprehensive understanding of the underlying dynamics. 
Measurements of production cross sections and searches for alternative decay channels would also be particularly informative. For instance, if the same resonance is observed in both the $J/\psi \phi$ and $J/\psi \omega$ systems---whose light-quark compositions differ---it would challenge a pure tetraquark interpretation. Likewise, if a structure appears in the $B^+ \to J/\psi \phi K^+$ decay but not in other production environments such as $\Upsilon(4S) \to J/\psi \phi$, it would raise important questions about its production mechanism and internal configuration. Comparative studies of production cross sections across different processes would also be helpful.
Additionally, the all-heavy $J/\psi J/\psi$ structures appear to offer some further insights into these questions~\cite{Zhu:2024swp}.

\section{Further investigation of the $J/\psi \phi$ system}

\subsection{Family of radial excited states of $1^{++}$ and the $X(4140)$ width?}
\indent

A triplet structure, $X(6600)$, $X(6900)$, and $X(7100)$, was reported in the $J/\psi J/\psi$ system~\cite{CMS:2023owd}. The mass differences between these states are approximately 200--300~MeV, and they appear to align along a Regge trajectory~\cite{Zhu:2024swp}. Moreover, the observed interference among these states implies identical $J^{PC}$. These observations suggest a possible family of radial excitations of a tetraquark~\cite{Zhu:2024swp}. 
Recently, CMS updated the triplet parameters with threefold higher precision~\cite{JJRun3PAS}. The widths of the triplet members are seen to decrease with a hypothesized radial quantum number~\cite{JJRun3PAS}.
% , following an approximately exponential trend~\cite{JJRun3PAS}. 
A systematic relationship among their widths further points to a familial relationship.

The $J/\psi \phi$ system shares similarities: both involve two neutral vector mesons with hidden flavor and $J^{PC}=1^{--}$, but they also display important differences. The $\phi$ meson consists of an $s\bar{s}$ pair, where the $s$ quark is lighter than the $c$ quark but heavier than the $u$ and $d$ quarks. Moreover, the final state involves two different flavors, $c$ and $s$, in contrast to the $J/\psi J/\psi$ system, which contains only the $c$ flavor. 
These distinctions may lead to not only different formation mechanisms, but also potentially different dynamical binding due to the differing symmetry properties of tetraquark wave functions for all-identical, versus mixed, quark flavors. 
In this work, following the implications presented in Ref.~\cite{Zhu:2024swp}, we explore the possibility that the $1^{++}$ triplet in the $J/\psi \phi$ system forms a family of excitations.

\begin{figure}[!htbp]
    \centering
    \includegraphics[width=0.48\linewidth]{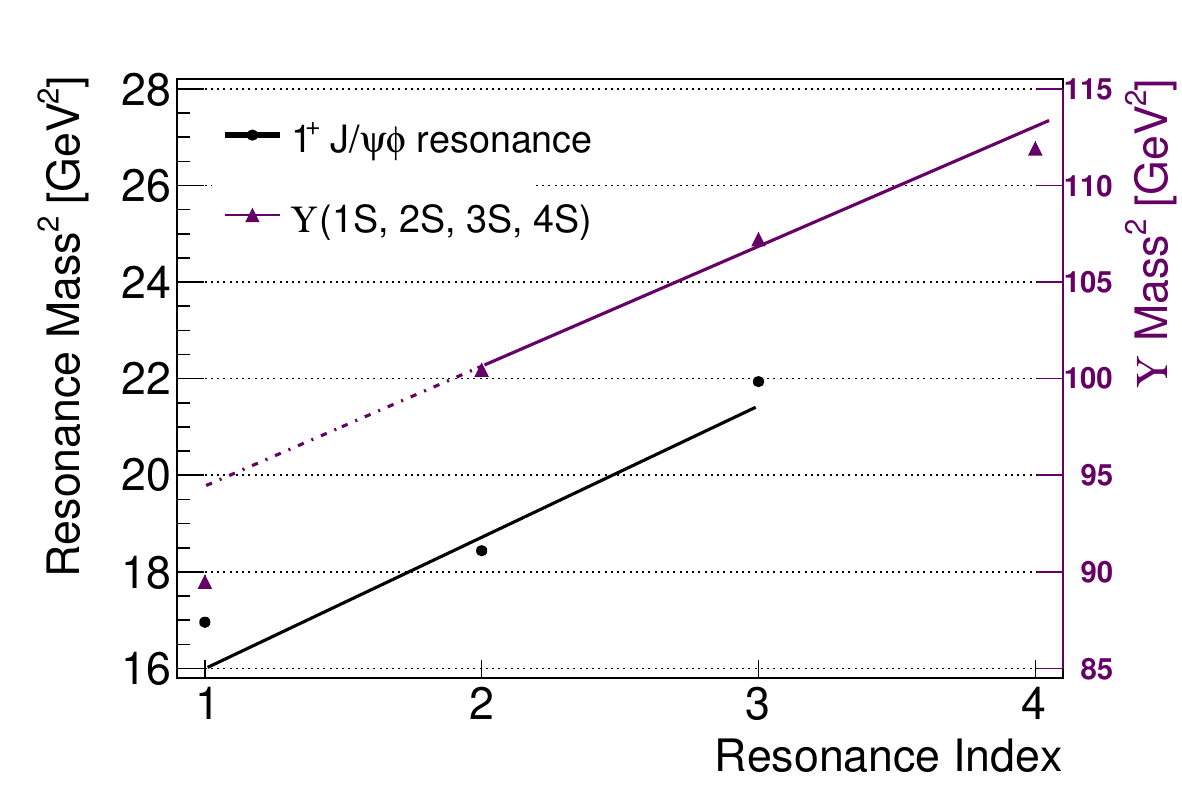}
    \includegraphics[width=0.46\linewidth]{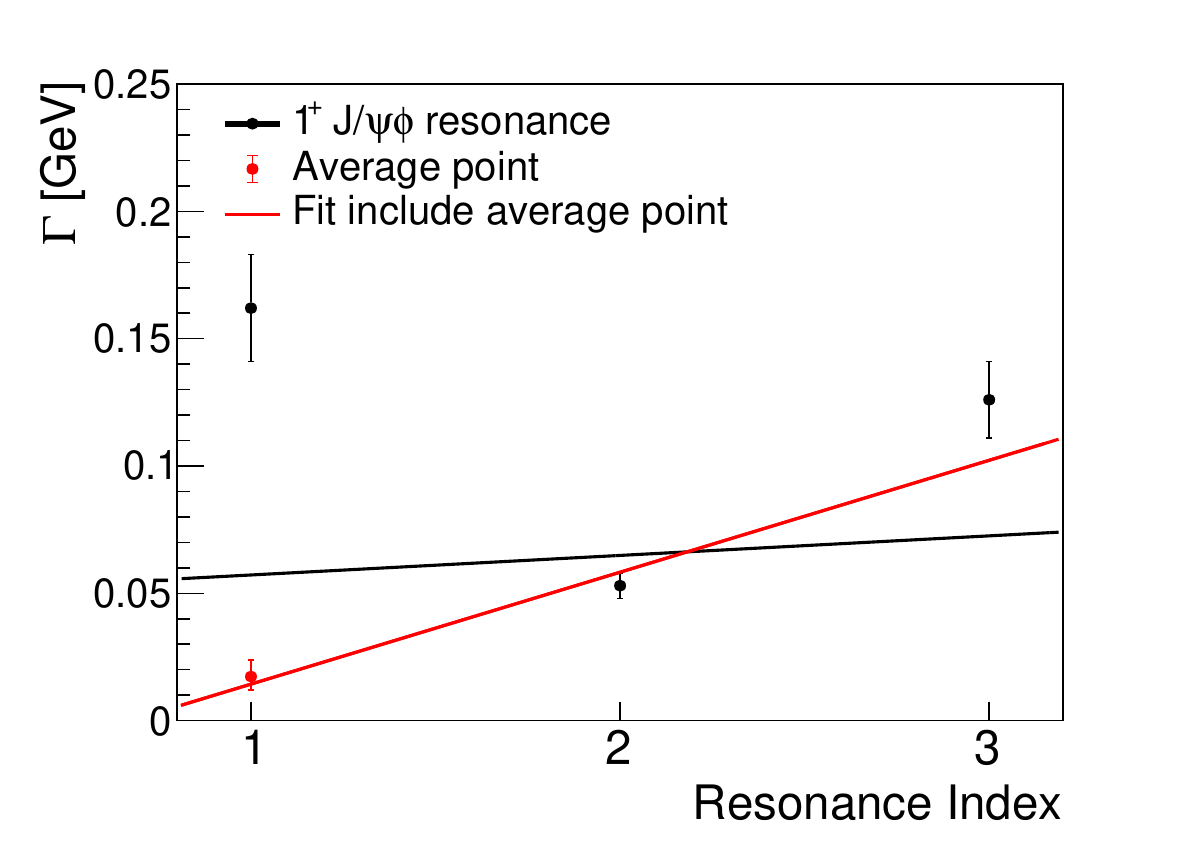}
    \caption{Left: The squared masses of the $1^{++}$ \(J/\psi \phi\) resonances versus hypothetical radial quantum numbers $n_{r} = 1$, $2$, $3$. The \(\Upsilon(nS)\) family is also shown with a linear fit of the $n_{r}=2$--$4$ excitations only (the apparent similarity of $J/\psi \phi$ resonances and $\Upsilon(nS)$ slopes is deceptive due to the different scales of the two axes). Right: Width of the $1^{++}$ \(J/\psi \phi\) resonances versus radial index. The red point shows the average width of $X(4140)$, calculated as $17.3^{+6.6}_{-5.3}$~MeV by combining results from CDF, CMS, and $\mathrm{D\textsl{\O}}$ (Tab.~\ref{tab:Comparison4140}), whereas the LHCb alternative is the corresponding black point. The black line refers to the linear fit for three black points, and the red line is the linear fit for the red point and two black points.}
    \label{fig:RegJpsiPhiMass}
\end{figure}

There are three resonances, $X(4140)$, $X(4274)$, and $X(4685)$, in the $J/\psi \phi$ system with the same $J^{PC}=1^{++}$ according to the LHCb results (Tab.~\ref{tab:jpsiphi_mass_spectrum}). The mass splittings among these states are relatively large (around 200~MeV), which is unlikely to occur due to spin-orbit splittings. The large mass splittings, together with their identical quantum numbers, suggest that they form a family of radial excitations. We plot their squared masses as a Regge trajectory with possible radial quantum numbers $n_r=1$, $2$, $3$ in Fig.~\ref{fig:RegJpsiPhiMass} (left). Traditionally, the Regge plot is used to present the squared mass as a function of the total angular momentum $J$, but it has also been applied to show the squared mass versus a radial quantum number $n_r$~\cite{Ebert:2011jc}. For comparison, the squared masses of the $\Upsilon$ family, which represents a well-established radially excited spectrum, are also plotted in Fig.~\ref{fig:RegJpsiPhiMass} (left). Both the $\Upsilon$ family and the triplet of $1^{+}$ $J/\psi \phi$ states are roughly linearly aligned with respect to a radial quantum number, supporting an interpretation of the triplet as a family of radially excited states.

Their widths are also shown in Fig.~\ref{fig:RegJpsiPhiMass} (right). A striking feature is the width of $X(4140)$. The result from LHCb is very different from the average width obtained from all other experiments. The two $X(4140)$ width values present two different potential trends in Fig.~\ref{fig:RegJpsiPhiMass} (right). There appears to be an increase in width with resonance mass for the non-LHCb $X(4140)$ widths, but a potential decrease followed by an increase for the LHCb $X(4140)$ width. The width trends in both cases are different from the $J/\psi J/\psi$ system, which exhibits a decrease~\cite{JJRun3PAS}, and the widths in the $J/\psi \phi$ system are also significantly narrower than those in the $J/\psi J/\psi$ system. 
Thus, it is important to obtain a full understanding of the width of $X(4140)$, as there might be an interesting width trend that can shed light on the nature of the $J/\psi \phi$ triplet. 
The result from LHCb may be more credible, as it is derived from a full-amplitude analysis, whereas others are obtained from one-dimensional fits. 
However, factors such as imperfect efficiency correction can also bias parameter extraction, particularly when the efficiency exhibits a complicated dependence on multiple kinematic variables, making a complete correction challenging.
We will provide an extended discussion of the discrepancies between the resonance results reported by LHCb and those obtained by other experiments, starting from a comparison of mass spectra.

\subsection{Comparison of $J/\psi \phi$ mass spectra from CMS and LHCb}
\indent

\begin{figure}[htbp!]
    \centering
    \includegraphics[width=0.4\linewidth]{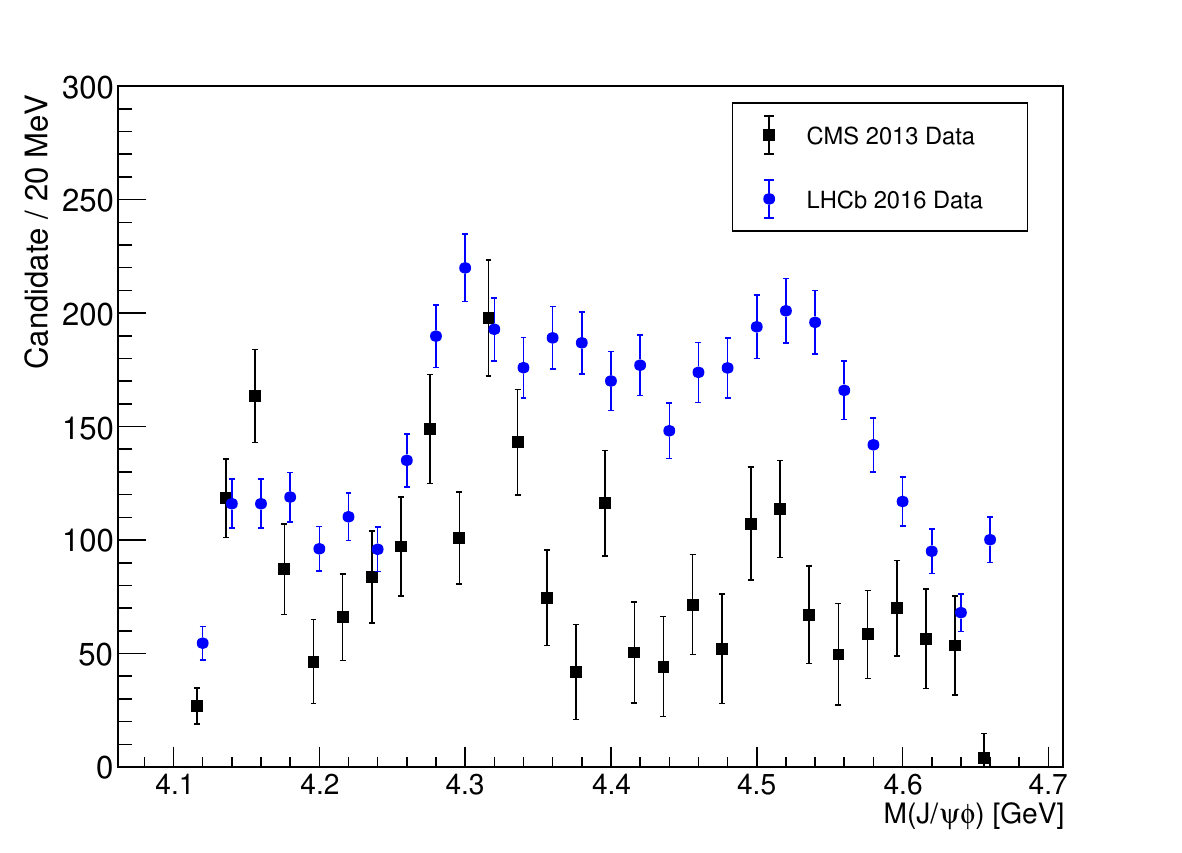}
    \includegraphics[width=0.4\linewidth]{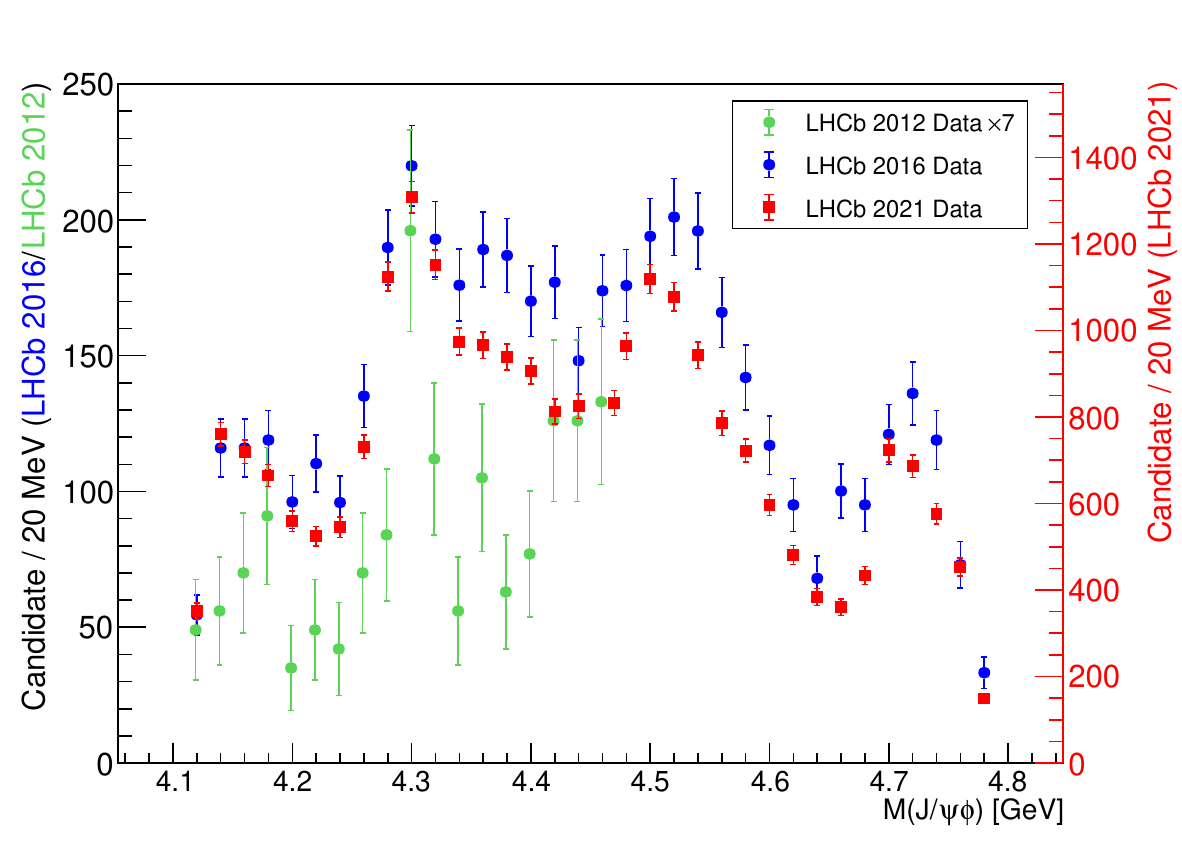}
    \includegraphics[width=0.4\linewidth]{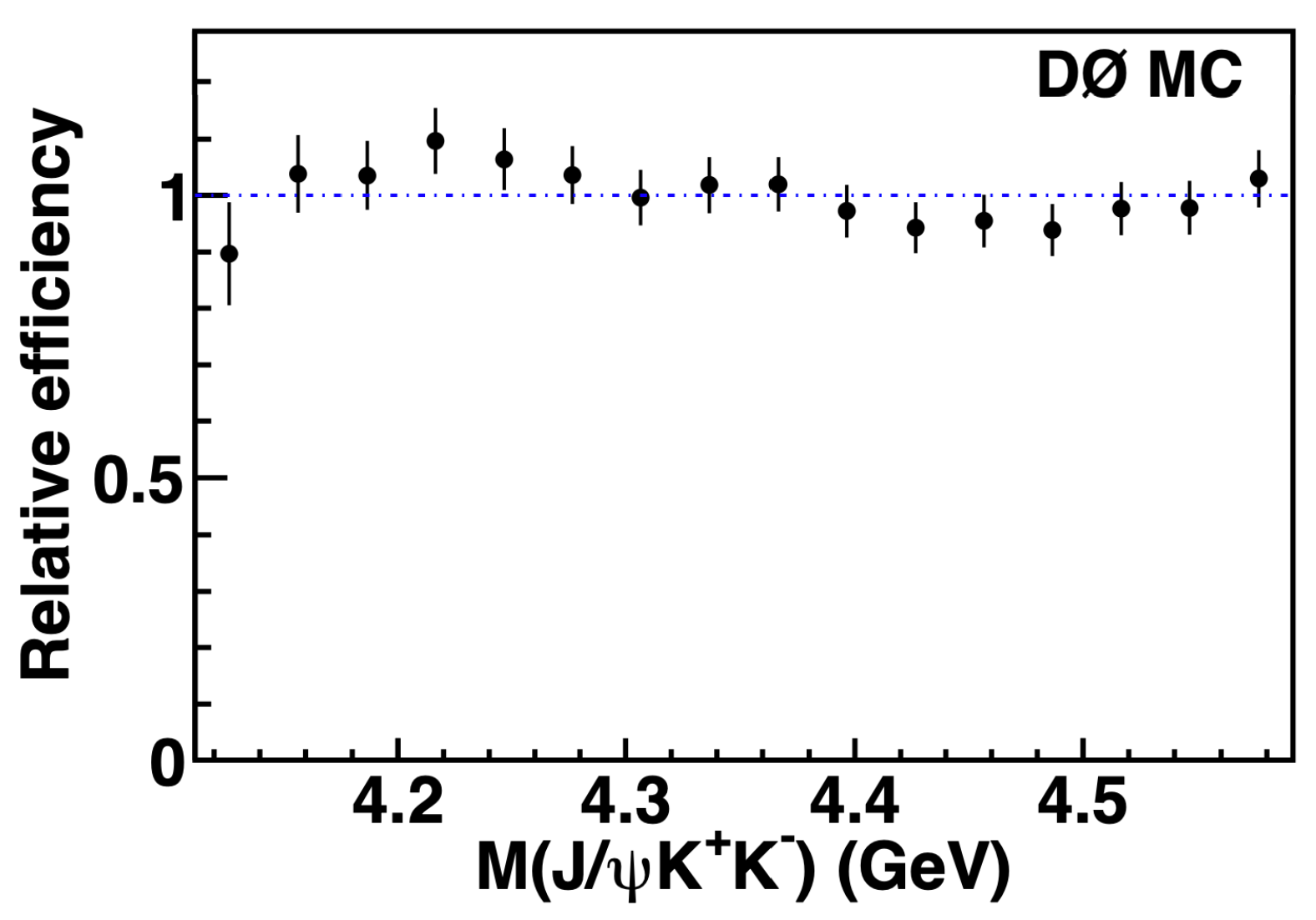}
    \includegraphics[width=0.3\linewidth]{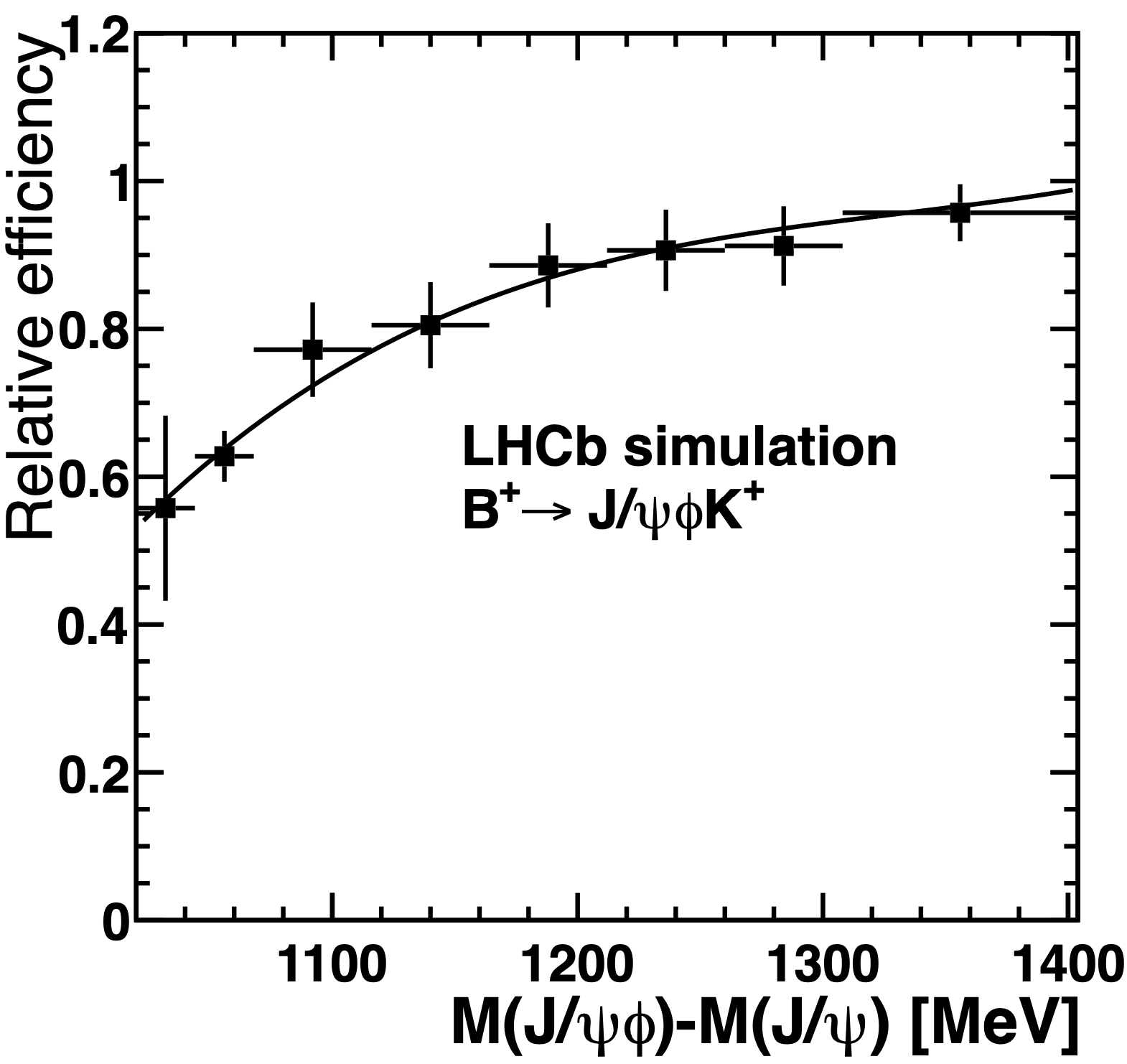}
    \caption{Upper left: The \( J/\psi \phi \) mass spectrum from LHCb 2016 results (Run 1)~\cite{LHCb:2016axx} and CMS 2013 results (part of Run 1)~\cite{CMS:2013jru}. Upper right: The comparisons in the mass spectra of the $J/\psi \phi$ system among the LHCb results from 2012 (part of Run 1)~\cite{LHCb:2012wyi}, 2016 (Run 1)~\cite{LHCb:2016axx}, and 2021 (Run 1 and Run 2)~\cite{LHCb:2021uow}. Lower left: Relative efficiency for $J/\psi K^+K^-$ in $\mathrm{D\textsl{\O}}$~\cite{D0:2013jvp}, where a relatively flat efficiency is observed with less than 10\% variations across bins. Lower right: The efficiency curve for \( \Delta M = M(J/\psi \phi) - M(J/\psi) \) (the $1.0<\Delta M<1.4$~GeV range approximately corresponds to $4.1<M(J/\psi \phi) < 4.5$~GeV) from LHCb 2012~\cite{LHCb:2012wyi}.}
    \label{fig:lhcbCompareCMS2}
\end{figure}

The limited statistics in the various experiments studying the $J/\psi \phi$ mass spectrum makes comparisons challenging. Relatively large statistics are only available in CMS and LHCb. The data for the $J/\psi \phi$ mass spectrum from CMS 2013 results (part of Run 1 data)~\cite{CMS:2013jru} and LHCb 2016 results (the whole Run 1 data)~\cite{LHCb:2016axx} are plotted together for easier comparison in Fig.~\ref{fig:lhcbCompareCMS2} (top left). The $B$ yield in the LHCb 2016 analysis is $4289\pm151$, compared to $2480\pm160$ in CMS, a difference of a factor of 1.7. The mass distributions from the two experiments differ substantially above about 4.3~GeV. The yield from LHCb is much higher than that from CMS in the high-mass region while that in the low-mass region are comparable. This raises the question of what factors may have caused these notable discrepancies.

The larger background in CMS data is a contributing factor, and another critical factor may be the differing efficiencies between the two experiments. 
The one-dimensional efficiency curve as a function of the $J/\psi \phi$ mass from $\mathrm{D\textsl{\O}}$ is shown in Fig.~\ref{fig:lhcbCompareCMS2} (bottom left), exhibiting relatively flat efficiency. 
No public one-dimensional efficiency curve is available from CDF or CMS, but CDF, CMS and $\mathrm{D\textsl{\O}}$ are all broadly similar central detectors. 
The efficiencies in CDF and CMS are slightly higher (about 20\%) near the low-mass threshold than in the high-mass region~\cite{private}. The efficiency curve for $\Delta M = M(J/\psi \phi) - M(J/\psi)$ [equivalent to $M(J/\psi \phi)$] for LHCb is depicted in Fig.~\ref{fig:lhcbCompareCMS2} (bottom right), showing a decrease of about 40\% near the threshold compared with the region around 4.5 ($\Delta M\approx 1.4$)~GeV, while no information is available around 4.7 ($\Delta M\approx 1.6$)~GeV. However, the trend of the efficiency curve from LHCb in Fig.~\ref{fig:lhcbCompareCMS2} (bottom right) shows a slight increase in the high-mass region, resulting in an efficiency variation exceeding 40\% across the full mass range. 
This significant efficiency difference between CMS and LHCb may partly explain the shape inconsistency (before efficiency correction) shown in Fig.~\ref{fig:lhcbCompareCMS2} (top left). This may also contribute to LHCb's higher sensitivity in the high-mass region, where two additional resonances were claimed, and its failure with their 2012 data~\cite{LHCb:2012wyi} to confirm the two states in the low-mass region reported by CDF and CMS.

A comparison of the LHCb $ J/\psi \phi$ mass spectrum from different years, 2012~\cite{LHCb:2012wyi}, 2016~\cite{LHCb:2016axx}, and 2021~\cite{LHCb:2021uow} is illustrated in Fig.~\ref{fig:lhcbCompareCMS2} (top right). The LHCb data from 2012 are statistically limited. However, they exhibit excess features in the regions around $X(4140)$ and $X(4274)$ after scaling up. Interestingly, the $ J/\psi \phi$ mass spectrum shapes from LHCb in 2016 and 2021 also show some inconsistencies across the mass spectrum. As there is no apparent change for the LHCb tracker, we do not expect efficiency changes from the LHCb tracker. Possible explanations include efficiency variations due to different triggers and event selections, as well as statistical fluctuations amplified by the rich resonance structure of the system.

\subsection{Detector efficiency: variations and potential impact}
\indent

Accurate efficiency corrections are essential in Dalitz plot analyses, as imperfect corrections can bias the extracted resonance parameters. In this section, we further examine the efficiency maps from different experiments to illustrate their differences, aiming to provide insight into how these variations may contribute to the discrepancies observed in the measured resonance parameters.

\begin{figure}[htbp!]
    \centering
    \includegraphics[width=0.32\linewidth]{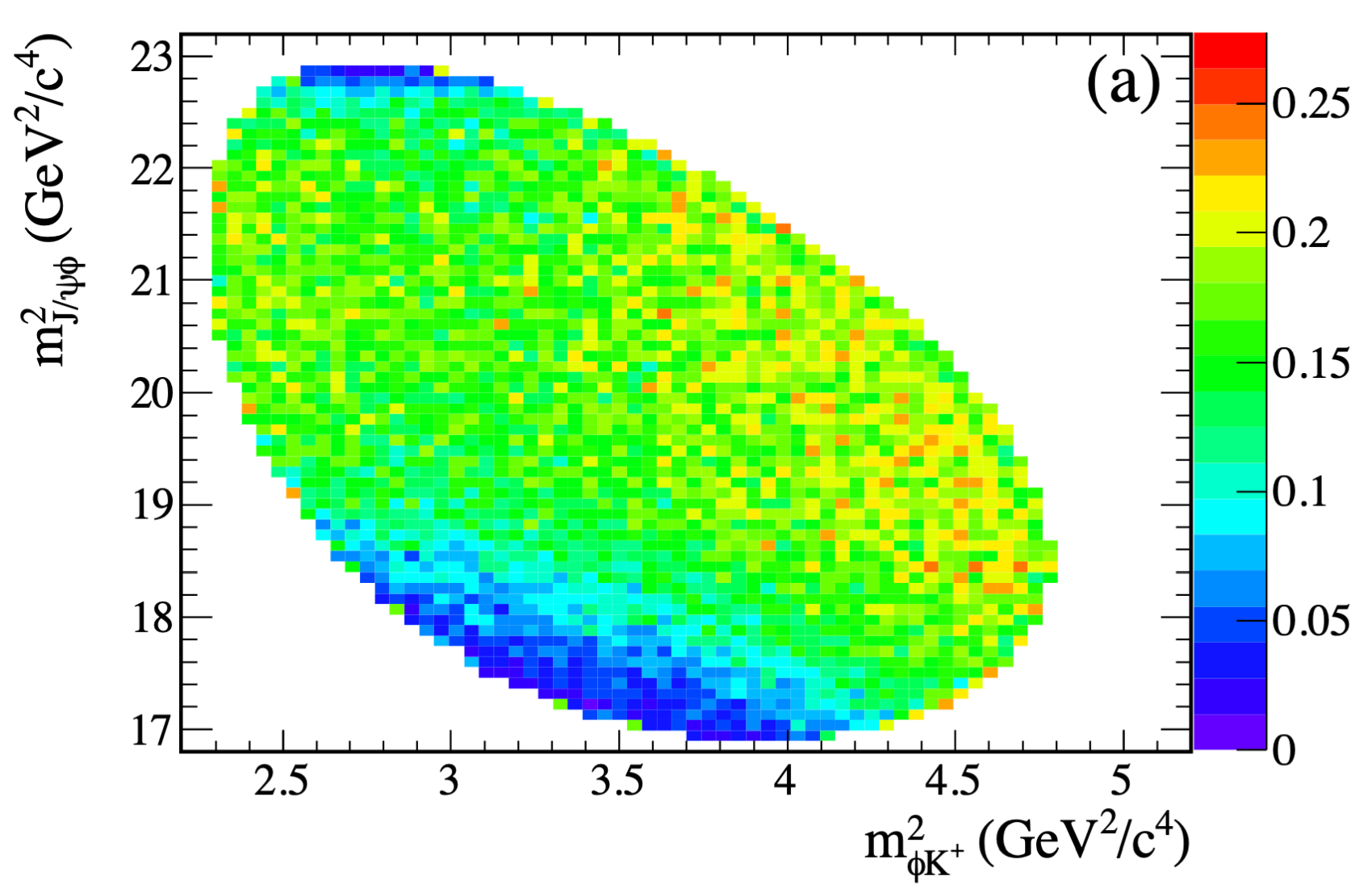}
    \includegraphics[width=0.32\linewidth]{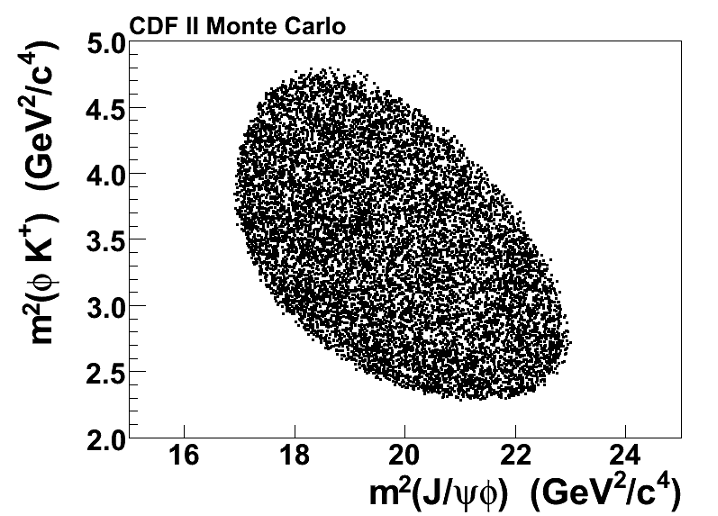}
    \includegraphics[width=0.32\linewidth]{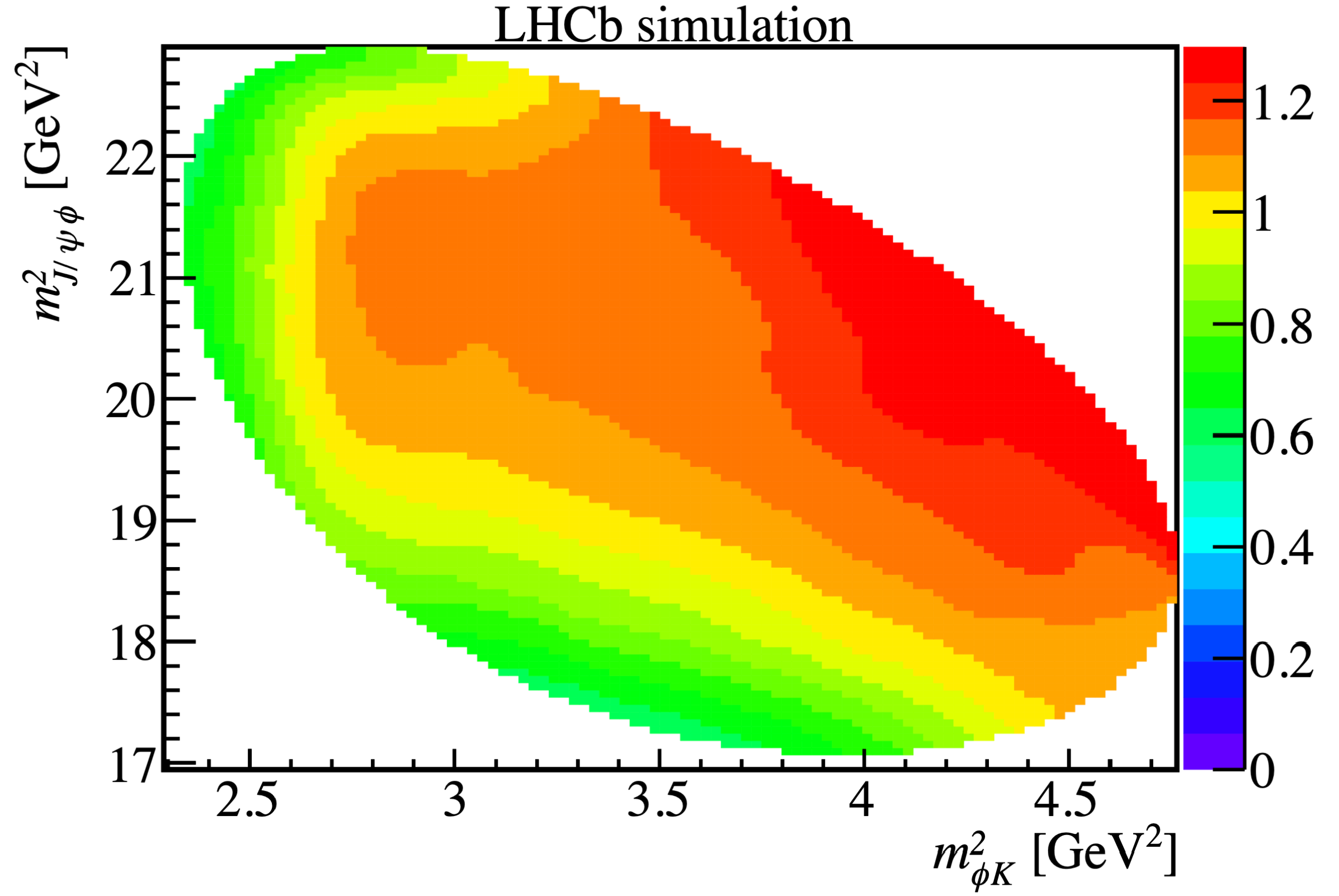}
    \caption{Two-dimensional efficiency distributions in the Dalitz plot for the decay $B^+ \to J/\psi \phi K^+$, shown as a function of $m^2_{J/\psi\phi}$ versus $m^2_{\phi K^+}$ in BaBar~\cite{BaBar:2014wwp} (left), CDF~\cite{Yi2009JETP} (middle) and LHCb~\cite{LHCb:2016nsl} (right).}
    \label{fig:effmap}
\end{figure}

Public two-dimensional efficiency maps in the Dalitz plane for $B^+ \to J/\psi \phi K^+$ are available from BaBar, CDF, and LHCb (Fig.~\ref{fig:effmap}). CDF shows a relatively flat distribution of MC phase-space events, while CMS, though lacking a public map, is expected to be similar due to its comparable detector design (see basic detector features in Tab.~\ref{tab:detector}). In contrast, BaBar and LHCb exhibit significant efficiency variations: BaBar is mostly flat but drops by $\sim25\%$ near the low/edge regions of $m^{2}(J/\psi \phi)$, while LHCb shows additional variation bands. Such differences can distort the raw mass spectra, and bias the extracted resonance parameters if efficiency corrections are not precise.

\begin{figure}[!htbp]
    \centering
    \includegraphics[width=0.5\linewidth]{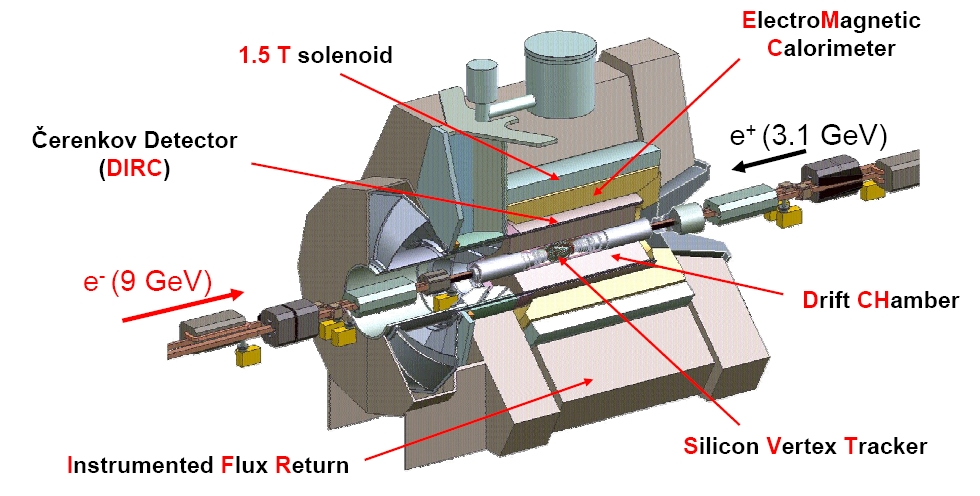}
    \includegraphics[width=0.32\linewidth]{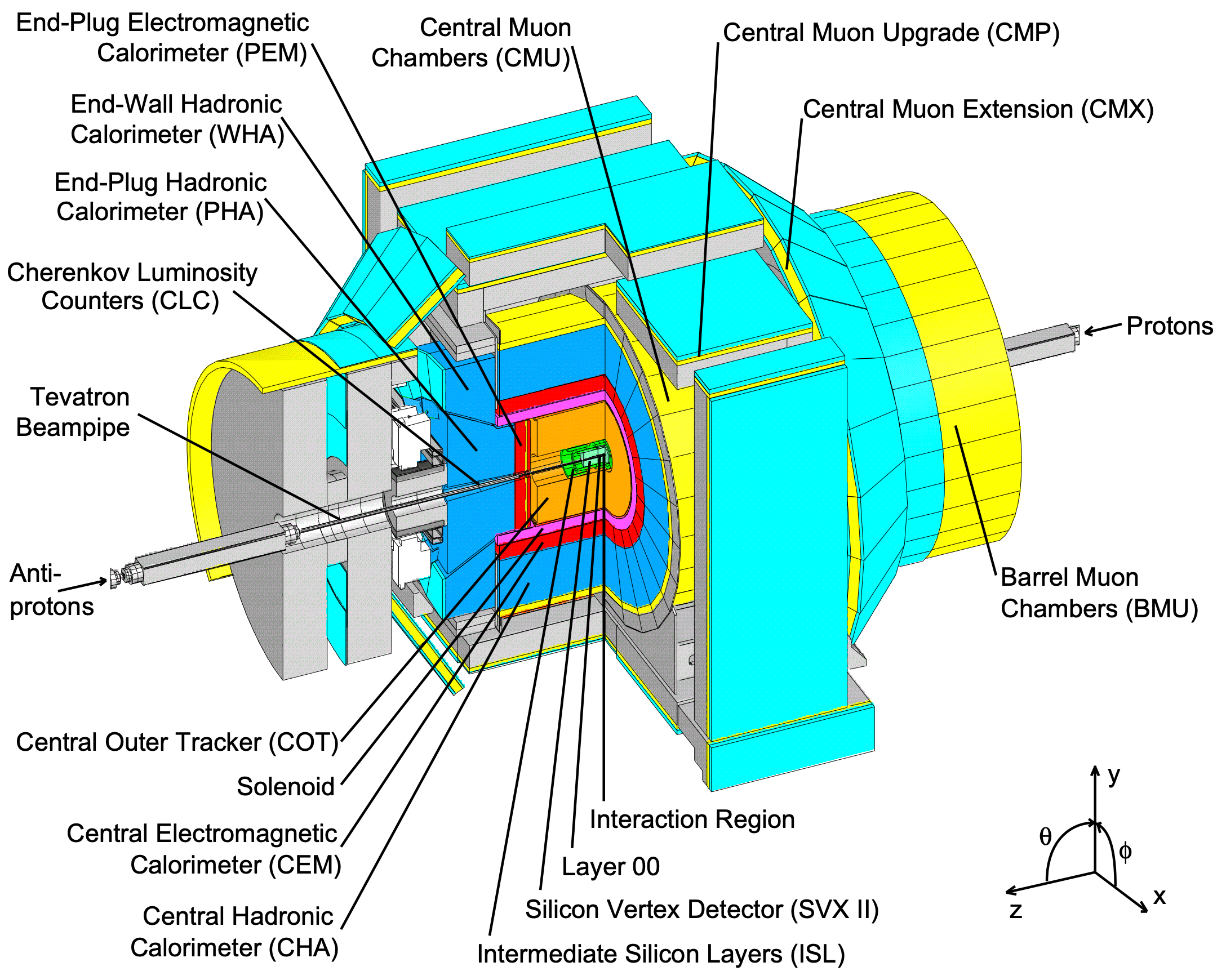}
    \vspace{0.5em}
    \includegraphics[width=0.48\linewidth]{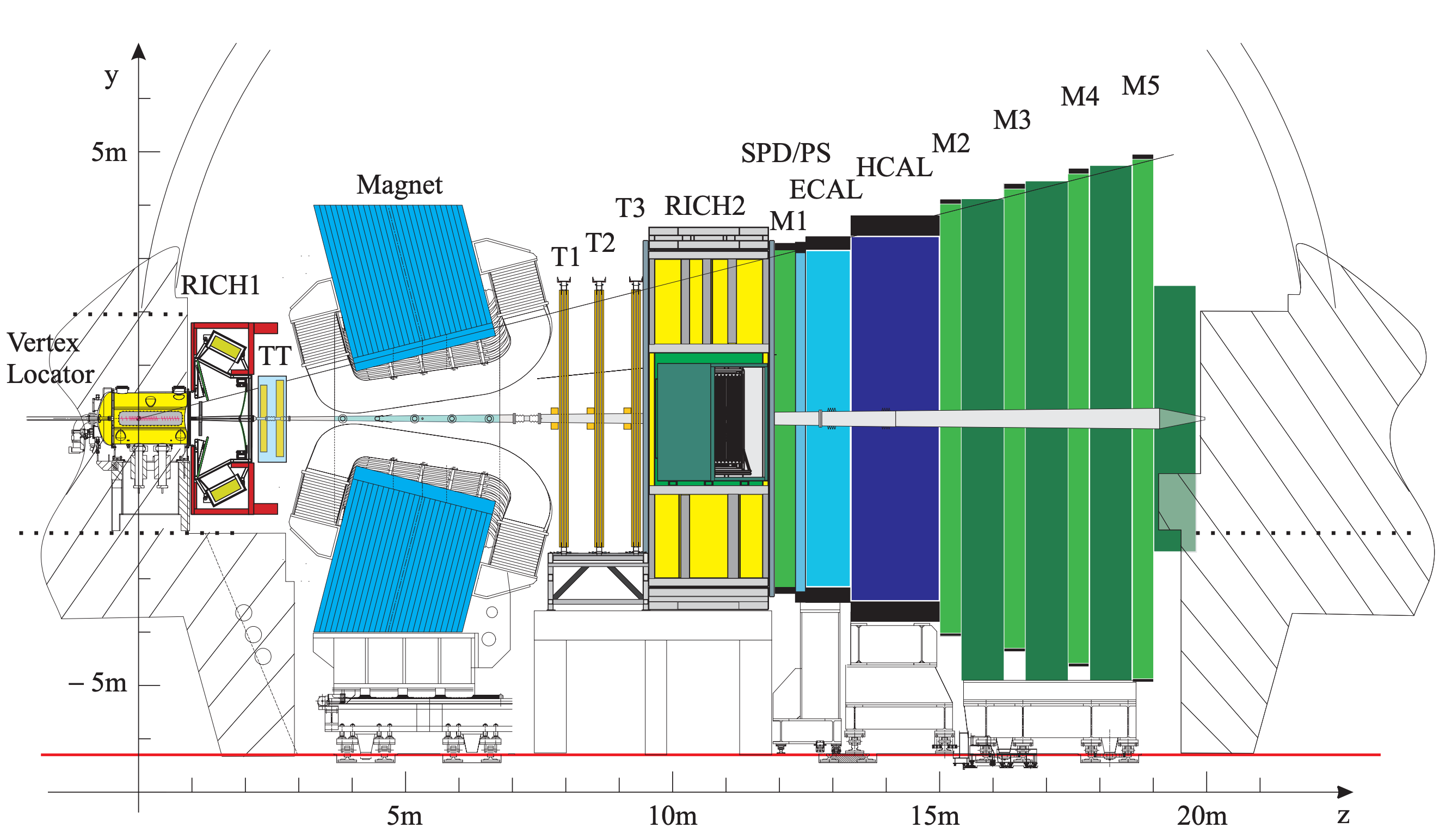}
    \caption{Top left: The BaBar detector longitudinal section~\cite{BaBar_detector_2001}; Top right: Cutaway isometric view of the CDF II detector~\cite{CDF_detector_1996,CDF:2013ezj}; Bottom: Cross section of the LHCb detector~\cite{LHCb_detector}.}
    \label{fig:detectors}
\end{figure}

Detector configurations help explain these behaviors (Fig.~\ref{fig:detectors} and Tab.~\ref{tab:detector}). BaBar/Belle ($e^+e^-$) reconstruct nearly at-rest $B$ mesons in the transverse plane; and near threshold $\phi$-decay kaons have little transverse momentum, reducing acceptance at the Dalitz edges. For central detectors (CDF, $\mathrm{D\textsl{\O}}$, CMS) at hadron colliders, $B$ mesons are highly boosted with near-$4\pi$ coverage, yielding flatter efficiencies. For LHCb, a forward detector, its forward acceptance yields strong longitudinal but modest transverse boosts, producing more pronounced Dalitz variations.

Revisiting the $X(4140)$ measurements (Tab.~\ref{tab:Comparison4140}), LHCb reports masses consistent with the non-LHCb average within $1\sigma$, but widths differ by more than $2\sigma$. Potential contributions from imperfect efficiency treatment cannot be excluded. Complementary measurements with flatter Dalitz-plane efficiency---from ATLAS, Belle II, and CMS---will be essential for cross-checks and improved precision.

\begin{table}[htbp!]
    \centering
    \caption{Comparison of the results for $X(4140)$ from different experiments. The average width from CDF, CMS and $\mathrm{D\textsl{\O}}$ is $17.3^{+6.6}_{-5.3}\,\mathrm{MeV}$. Using the formula $\frac{\Delta x}{\sqrt{\sigma_{1}^{2}+\sigma_{2}^{2}}}$, the width measurement from LHCb 2016 differs from this average by $^{+2.2}_{-2.6}\sigma$ and that from LHCb 2021 differs by $^{+4.4}_{-2.7}\sigma$ (the asymmetry arises from using the lower or upper uncertainties in the denominator). The shifts in $X(4140)$ mass and width between the LHCb 2016 and 2021 results may arise from the fact that the LHCb 2021 analysis introduces additional $J/\psi \phi$ structures in the fit.}
    \begin{tabular}{c|c|c}
    \hline
    \hline
        Detector & $M[(X(4140)]\, [\mathrm{MeV}]$ & $\Gamma[X(4140)]\, [\mathrm{MeV}]$ \\
    \hline
        CDF~\cite{CDF:2011pep} & $4143.4^{+2.9}_{-3.0}\pm0.6$ & $15.3^{+10.4}_{-6.1}\pm2.5$ \\
        CMS~\cite{CMS:2013jru} & $4148.0\pm2.4\pm6.3$ & $28.0^{+15}_{-11}\pm19$ \\
        $\mathrm{D\textsl{\O}}$ 2013~\cite{D0:2013jvp} & $4159.0\pm4.3\pm6.6$ & $19.9\pm12.6^{+1.0}_{-8.0}$ \\
        $\mathrm{D\textsl{\O}}$ 2015~\cite{D0:2015nxw} & $4152.5\pm1.7^{+6.2}_{-5.4}$ & $16.3\pm5.6\pm11.4$ \\
        \hline
        LHCb 2016 (Run 1 data)~\cite{LHCb:2016axx} & $4146.5\pm4.5^{+4.6}_{-2.8}$ & $83\pm21^{+21}_{-14}$ \\
        LHCb 2021 (Run 1 and Run 2 data)~\cite{LHCb:2021uow} & $4118\pm11^{+19}_{-36}$ & $162\pm21^{+24}_{-49}$ \\
        \hline
        Average excluding LHCb & $4146.6\pm 2.4$ & $17.3^{+6.6}_{-5.3}$ \\
    \hline
    \hline
    \end{tabular}
    \label{tab:Comparison4140}
\end{table}

\subsection{A potential high-spin state around 4.35~GeV in the $J/\psi \phi$ mass spectrum?}
\indent

Having discussed the detector-efficiency variations and their potential impact on the extracted resonance parameters, we now proceed to investigate a potential resonance in the $J/\psi\phi$ channel.
There is evidence of an excess at $4350.6^{+4.6}_{-5.1}\pm0.7$~MeV with a width of $13.9^{+18}_{-9}\pm4$~MeV reported by Belle through the double-photon process to $J/\psi \phi$ (Fig.~\ref{fig:Belle_BESIII_noY4140}, left)~\cite{Belle:2009rkh}. The spin of this 4.35~GeV excess could be 0 or 2 since a massive spin-1 particle cannot decay into two identical massless spin-1 bosons via a two-photon process according to the Landau–Yang theorem~\cite{Landau:1948kw,Yang:1950rg}. Several tetraquarks have been predicted around 4.3~GeV~\cite{Wu:2016gas}. However, such a resonance is not claimed in $B\rightarrow J/\psi \phi K$ by LHCb~\cite{LHCb:2016axx,LHCb:2021uow}. On the other hand, a notable shoulder between 4.32 and 4.39~GeV is observed (Fig.~\ref{fig:LHCb2021_Jpsiphi_massSpetrum}, middle) in the \( J/\psi \phi \) mass spectrum, while the LHCb 2016 model fails to capture features in this region (Fig.~\ref{fig:LHCb2021_Jpsiphi_massSpetrum}, bottom middle). The updated 2021 model provides a better description by introducing additional components (Fig.~\ref{fig:LHCb2021_Jpsiphi_massSpetrum}, top middle). In the updated model, the shoulder is described by the superposition of the tails from $X(4274)$, $X(4500)$, and other $\phi K$ and $J/\psi K$ resonances, without requiring any extra $J/\psi \phi$ resonance that peaks in this region with a significance above $5\sigma$ (Fig.~\ref{fig:LHCb2021_Jpsiphi_massSpetrum}, top). Although the updated model captures data more accurately, it comes at the cost of invoking a larger set of states, many of which remain to be confirmed. Given the complexity of the system, we explore alternative descriptions for the shoulder.

An alternative explanation for this shoulder in LHCb data is the contribution of a resonance around 4.35~GeV in the $J/\psi \phi$ mass spectrum, motivated by the Belle finding~\cite{Belle:2009rkh}. The shoulder-like feature could be related to production suppression if this state carries high spin~\cite{spin2discuss}.
Since both the $B$ and $K$ mesons are spin-0, producing a spin-2 (or higher) particle in association with a kaon requires orbital angular momentum $L=2$ to conserve the total angular momentum. This introduces a centrifugal barrier, suppressing production, so the $J/\psi \phi$ invariant-mass spectrum around 4.35~GeV can appear as a shoulder or a small enhancement rather than a distinct peak. 
Another interpretation is that this structure is a hybrid which is suppressed in B decay, since B decay does not produce an environment rich in gluons~\cite{private_hybrid}.

To investigate this possibility, we generated a toy sample based on the LHCb 2021 model to reproduce the mass and angular distributions observed in the combined LHCb Run 1 and Run 2 data. A multidimensional amplitude fit was then performed on this sample to estimate the significance of a potential structure near 4.35~GeV in the $J/\psi \phi$ mass spectrum. This structure was provisionally named $X(4350)$. To fit the toy sample, we considered two fit models: a full-set model and a reduced model. The full-set model followed LHCb 2021 parameterization (Fig.~\ref{fig:LHCb2021_Jpsiphi_massSpetrum}, bottom), which described LHCb data well, with the additional $X(4350)$ component. The reduced model only includes $X(4350)$, $X(4140)$, $X(4274)$, and two other with the most significant $J/\psi \phi$ contributions, while keeping the $J/\psi K$ and $\phi K$ states identical to the full-set model, since most $J/\psi \phi$ structures beyond $X(4140)$ and $X(4274)$ remain unconfirmed by other experiments.

For a given fit model, the significance of $X(4350)$ was calculated from the log-likelihood difference ($\Delta \mathrm{ln}L$) between fits with and without this component. In principle, we could fix the mass and width of $X(4350)$ to Belle's measured values. However, due to Belle’s limited statistics, the mass and width values are not very well determined. We instead scanned several fixed masses (4320, 4340, 4360~MeV) and widths (15, 50, 100~MeV), under spin-parity hypotheses $0^{+}$, $0^{-}$, $2^{+}$, $2^{-}$, as well as $1^{+}$ and $1^{-}$ for completeness.

To reduce computational complexity, fitting time, and potential instability, the parameters related to the $J/\psi K$ and $\phi K$ structures, as well as the mass and width of the known $J/\psi \phi$ structures, were fixed to the values obtained by LHCb, while the amplitudes of the $J/\psi \phi$ structures were allowed to float freely. The resulting significances of this potential $X(4350)$ for each spin-parity hypothesis with the full-set model are summarized in Tab.~\ref{tab:7X_X4350}. A hypothesized spin $2^+$ or $2^-$ with a mass of 4.32~GeV and a width of 100~MeV shows a significance of 3.7$\sigma$ in this test.

\begin{figure}[htbp!]
    \centering
    \includegraphics[width=0.8\linewidth]{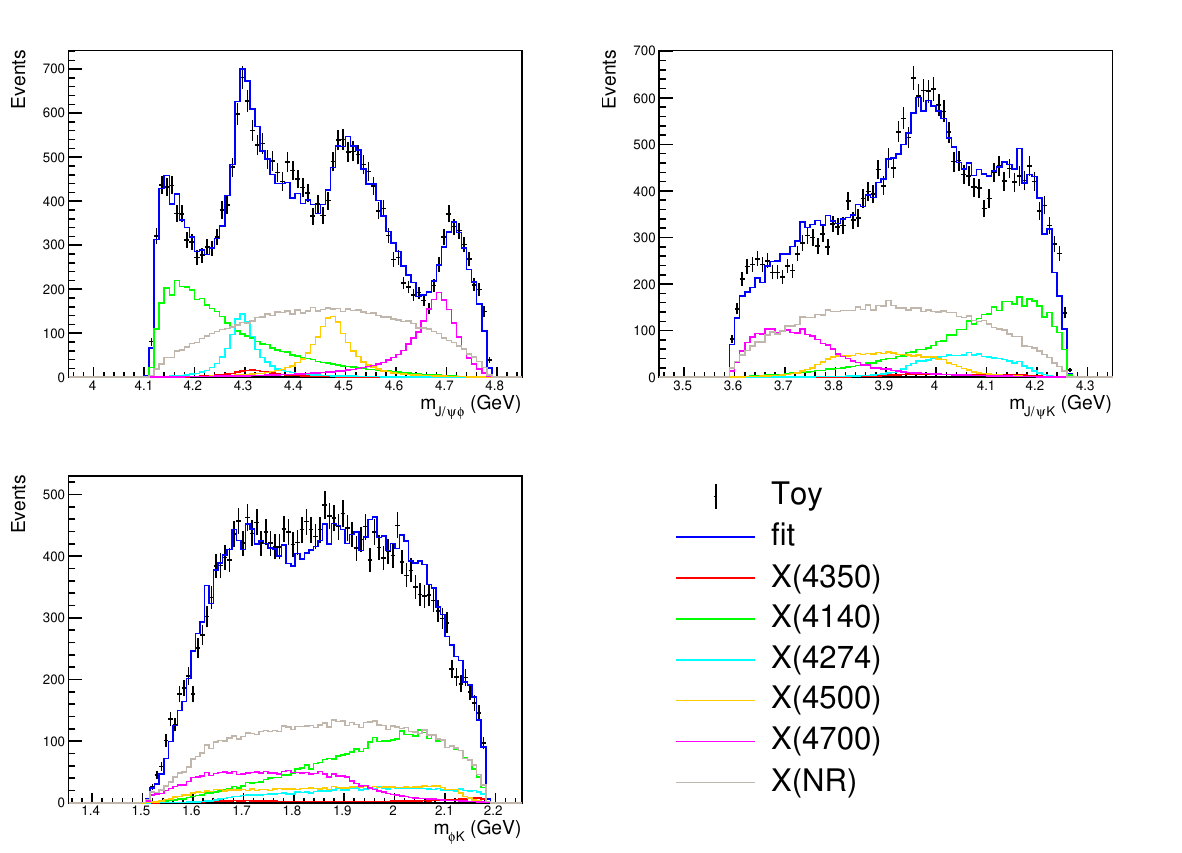}
    \caption{The $J/\psi \phi$, $\phi K$, and $J/\psi K$ mass distributions along with the fit models for the most significant case of $X(4350)$ ($2^{-}$) in Tab.~\ref{tab:4X_X4350}. The toy sample is presented by points, featuring similar two-body invariant-mass and angular distributions observed in the combined LHCb Run 1 and Run 2 data.}
    \label{fig:4X_X4350}
\end{figure}

Since many of the $J/\psi \phi$ structures reported by LHCb have not been identified by other experiments, apart from $X(4140)$ and $X(4274)$, we tested an alternative fit model with fewer $J/\psi \phi$ components. In this reduced fit model (reduced-set model), the $J/\psi \phi$ components included only $X(4140)$ and $X(4274)$, the two most significant states $X(4500)$ and $X(4700)$, and the potential $X(4350)$. The same toy sample was analyzed using this reduced model, and the significance of $X(4350)$ is again estimated from the $\Delta \mathrm{ln}L$ between fits with and without this component. Significances of $X(4350)$ under different masses, widths, and spin-parity hypotheses are summarized in Tab.~\ref{tab:4X_X4350}. The most significant case (11.9$\sigma$) corresponded to spin $2^-$, at 4.32~GeV mass and 100~MeV width. The mass distributions and fit models for this case are presented in Fig.~\ref{fig:4X_X4350}. 
Although $X(4350)$ is not visually prominent in Fig.~\ref{fig:4X_X4350}, the fit yields a relatively large significance, which may originate from the fact that its parameters were fixed in the test, and that this component helps to describe certain features of data across multiple dimensions.
The significances for $X(4350)$ from Tab.~\ref{tab:4X_X4350} tend to be higher than those in Fig.~\ref{tab:7X_X4350}, suggesting that $X(4350)$ may share characteristics with the newly introduced structures in the LHCb 2021 model, thereby motivating an investigation of $X(4350)$ in a more transparent and simplified picture.

According to Tabs.~\ref{tab:7X_X4350} and \ref{tab:4X_X4350}, spin-2 hypotheses generally yield higher significances than spin-0. These tests point to the possibility of a potential high-spin $J/\psi \phi$ structure near 4.35~GeV. Its absence in the LHCb analyses may be attributed to the limited statistics and the suppressed production of this state. With larger statistics in the future, the sensitivity to such a state should improve. Independent information from other experiments could also help to clarify whether this state can be identified in $B$ decays, and potentially confirm Belle's report~\cite{Belle:2009rkh}.

\renewcommand{\arraystretch}{1.3}
\begin{table}[htbp!]
\scriptsize
    \centering
    \caption{Full-set model [includes all components in the 2021 LHCb model (Fig.~\ref{fig:LHCb2021_Jpsiphi_massSpetrum}, top) with additional structure $X(4350)$]: the significance of $X(4350)$ for each spin-parity hypothesis based on a MC toy sample. We consider the $X(4350)$'s masses 4320, 4340, and 4360~MeV, each tested with widths 15, 50, and 100~MeV. The significance [$\sigma$] is determined from the likelihood difference $\Delta\ln L$ between the fits with and without the $X(4350)$ component, and is approximated as $\sqrt{-2\Delta\ln L}$ when considering the degrees of freedom is 1.}
    \begin{tabular}{c|c|c|c|c|c|c|c|c|c}
    \hline
    \hline
         $J^{P}$ & $M=4320$ & $M=4340$ & $M=4360$ & $M=4320$ & $M=4340$ & $M=4360$ & $M=4320$ & $M=4340$ & $M=4360$ \\
         & $\Gamma=15$ & $\Gamma=15$ & $\Gamma=15$ & $\Gamma=50$ &  $\Gamma=50$ &  $\Gamma=50$ &  $\Gamma=100$ &  $\Gamma=100$ &  $\Gamma=100$ \\
         \hline 
$0^{+}$ & 2.0 & 1.4 & 0.5 & 2.3 & 1.9 & 1.4 & 2.3 & 2.1 & 1.9 \\
$0^{-}$ & 1.5 & 1.4 & 1.2 & 1.4 & 1.5 & 1.2 & 1.2 & 1.1 & 1.0 \\
$2^{+}$ & 2.7 & 3.0 & 2.2 & 3.4 & 3.3 & 3.0 & 3.7 & 3.6 & 3.4 \\
$2^{-}$ & 1.7 & 2.3 & 1.7 & 2.6 & 2.3 & 2.8 & 3.7 & 3.4 & 3.5 \\
$1^{+}$ & 2.4 & 1.5 & 2.2 & 2.0 & 1.9 & 2.1 & 1.7 & 1.6 & 1.8 \\
$1^{-}$ & 2.3 & 2.8 & 1.8 & 2.8 & 2.9 & 2.7 & 3.0 & 3.0 & 3.1 \\
        \hline
        \hline
    \end{tabular}
    \label{tab:7X_X4350}
\end{table}

\renewcommand{\arraystretch}{1.3}
\begin{table}[htbp]
\scriptsize
    \centering
    \caption{Reduced-set model [only four significant states $X(4140)$, $X(4274)$, $X(4500)$, and $X(4700)$ for $J/\psi \phi$ components are kept]: the significance of $X(4350)$ for each spin-parity hypothesis based on a MC toy sample. We consider the $X(4350)$'s masses 4320, 4340, and 4360~MeV, each tested with widths 15, 50, and 100~MeV.}
    \begin{tabular}{c|c|c|c|c|c|c|c|c|c}
    \hline
    \hline
         $J^{P}$ & $M=4320$ & $M=4340$ & $M=4360$ & $M=4320$ & $M=4340$ & $M=4360$ & $M=4320$ & $M=4340$ & $M=4360$ \\
         & $\Gamma=15$ & $\Gamma=15$ & $\Gamma=15$ & $\Gamma=50$ &  $\Gamma=50$ &  $\Gamma=50$ &  $\Gamma=100$ &  $\Gamma=100$ &  $\Gamma=100$ \\
         \hline 
$0^{+}$ & 3.5 & 2.6 & 1.3 & 3.5 & 3.3 & 2.6 & 3.4 & 3.3 & 3.1 \\
$0^{-}$ & 2.0 & 3.3 & 3.4 & 4.8 & 5.3 & 5.5 & 6.2 & 6.4 & 6.5 \\
$2^{+}$ & 3.5 & 3.2 & 2.8 & 4.7 & 4.4 & 4.3 & 5.8 & 5.5 & 5.4 \\
$2^{-}$ & 4.9 & 3.5 & 3.9 & 8.6 & 7.3 & 6.9 & 11.9 & 10.8 & 10.1 \\
$1^{+}$ & 3.7 & 2.7 & 4.3 & 5.7 & 5.5 & 6.0 & 7.1 & 7.3 & 7.7 \\
$1^{-}$ & 3.8 & 5.2 & 5.3 & 6.2 & 7.3 & 8.0 & 7.8 & 8.8 & 9.6 \\
        \hline
        \hline
    \end{tabular}
    \label{tab:4X_X4350}
\end{table}

\section{Prospects}
\indent

As mentioned above, there are near-threshold enhancements in vector–vector systems like $J/\psi \phi$ and $J/\psi J/\psi$. Interestingly, other vector–vector final states involving light quarks, studied at BES and other experiments, also display vector-vector enhancements. These near-threshold enhancements all seem to appear when both vectors have zero isospin, such as in $J/\psi \omega$~\cite{BaBar:2007vxr}, $\omega \phi$~\cite{BES:2006vdb}, and $\phi \phi$~\cite{BES:2008ecl,BESIII:2016qzq}. 
The same behavior appears in vector–vector systems without light quark involvement, such as $J/\psi \phi$ and $J/\psi J/\psi$. 
However, clear threshold enhancements seem not to be observed if one of the vectors has isospin 1, for instance, $\rho \phi$ or $\rho \omega$~\cite{BES:2007ahd}. 
Comparing these systems may provide insight into broader questions in exotic hadron spectroscopy. For instance, the near-threshold enhancements observed in these vector–vector systems raise the question of whether such phenomena represent a general pattern or accidental occurrences, and what theoretical mechanisms could account for them. 
If this pattern represents a more general "rule", we would expect near-threshold enhancements in the $J/\psi \Upsilon$ and $\Upsilon \Upsilon$ systems as well, since both $J/\psi$ and $\Upsilon$ have zero isospin and are narrow resonances.

We reiterate the point made in Ref.~\cite{YI:2013iok}: it is essential to systematically study vector–vector structures composed entirely of $c$ and $b$ quarks. Although the $s$ quark is not considered heavy, it is approximately 40 times heavier than the $u$ and $d$ light quarks, and thus serves as an intermediate testing ground. 
All $J/\psi \phi$, $J/\psi J/\psi$, $J/\psi \Upsilon$, and $\Upsilon \Upsilon$ structures contain heavy or quasi-heavy quarks, where their constituents are not highly relativistic, and the strong coupling constant $\alpha_s(Q^2)$ runs to smaller values at higher momentum scales associated with heavy-quark masses. Consequently, these systems provide theoretically cleaner environments in which QCD calculations are more controlled and reliable.
Heavy tetraquark states typically exhibit larger mass differences than traditional mesons or baryons, unlike light-quark tetraquark candidates. For instance, the $J/\psi J/\psi$ triplet lies about 3~GeV above the $J/\psi$ meson, and it is implausible to be an excited charmonium state, making it easier to distinguish from conventional mesons. Furthermore, final-state vectors composed only of heavy, or quasi-heavy, quarks have no isospin, which prevents binding through pion exchange.
Since both vectors are electrically neutral, photon exchange is also unlikely to contribute to binding. Among these systems, $J/\psi \phi$ and $J/\psi \Upsilon$ involve two different flavors, whereas the others are composed of identical flavors. Due to wave function symmetry requirements, identical versus non-identical quark flavors can potentially lead to different tetraquark configurations, which require experimental exploration to clarify.

Currently, the $J/\psi J/\psi$ triplet exhibits characteristics that show a preference for being a family of spin 2, radially excited all-charm tetraquarks composed of spin-1 diquarks~\cite{Zhu:2024swp,CMS:2025fpt,JJRun3PAS}. However, the situation for the $J/\psi \phi$ system remains unclear, and near-threshold structures in both the $J/\psi \Upsilon$ and $\Upsilon \Upsilon$ systems are, so far, unobserved. A systematic exploration of the $J/\psi \Upsilon$ and $\Upsilon \Upsilon$ systems, along with further studies of the $J/\psi \omega$, $\phi \phi$, $\rho \omega$, and $\rho \phi$ systems, will provide valuable experimental results. This will first help construct tetraquark models for all-heavy and quasi-heavy systems, which are theoretically simpler, and can later be extended to more complex tetraquark systems that include light quarks or are composed entirely of light quarks. Such efforts are expected to benefit from the capabilities of current high-performance experiments.

\section{Summary}
\indent

Inspired by the 2009 report of $X(4140)$ by CDF~\cite{CDF:2009jgo}, extensive investigations have been performed in the $J/\psi \phi$ and $J/\psi K$ systems. 
The article provides a brief overview of studies of the $J/\psi \phi$ system prior to 2013, but focuses primarily on experimental developments related to the $J/\psi \phi$ and $J/\psi K$ structures from 2013 to the present. 
Among the observed structures, the $1^{++}$ triplet in the $J/\psi \phi$ system is particularly striking: the three states exhibit large mass splittings, and their mass-squared approximately align linearly with a hypothetical radial quantum number, suggesting a pattern reminiscent of radial excitations. At the same time, experimental challenges remain. Some of these reported structures are not confirmed by other experiments, and the width of $X(4140)$ measured by LHCb shows inconsistency with earlier results. Possible reasons for the discrepancies among different measurements are discussed. A notable one is the substantial variation in the efficiency of detecting $J/\psi \phi$ final states across experiments, and nonuniform efficiency with imperfect correction can bias measured resonance parameters. This underlines the importance of experiments with relatively flat efficiency. Another open question concerns a potential resonance around 4.35~GeV in the $J/\psi \phi$ spectrum from $B$ decay, motivated by Belle’s observation in two-photon processes. There is a notable shoulder around 4.35~GeV in the $J/\psi \phi$ spectrum from LHCb data and its possible interpretation as a high-spin resonance remains under investigation. Taken together, these developments illustrate both the richness and the complexity of the $J/\psi \phi$ system, highlighting the need for further systematic exploration across multiple experiments.

\section{Acknowledgments}
This work is partially supported by the Natural Science Foundation of China under Grants No. 12075123, No. 12061141002, and No. 12535004, as well as the Ministry of Science and Technology of China under Grants No. 2023YFA1605804 and No. 2024YFA1610501.

\bibliographystyle{unsrt}
\bibliography{reference}
\end{document}